\documentclass[%
 a4paper,reprint, prl,
 amsmath,amssymb,
 showkeys,
]{revtex4-2}

\usepackage{bm,graphicx,xcolor}
\usepackage[activate={true,nocompatibility},final,tracking=true,kerning=true,spacing=true,factor=1100,stretch=10,shrink=10]{microtype}
\usepackage{hyperref}
\definecolor{hgwrot}{HTML}{da121a}
\hypersetup{colorlinks=true,linkcolor=hgwrot,citecolor=hgwrot,filecolor=hgwrot,urlcolor=hgwrot}

\let\epsilon\varepsilon
\let\theta\vartheta
\let\phi\varphi
\let\vec\bm

\begin{document}

\title{Geometry-Induced Transport of Self-Aligning Chiral Bristlebots}

\author{Timo Wagner}
\author{Michael Himpel}
\author{Thomas Ihle}
\author{Horst-Holger Boltz}
\email{horst-holger.boltz@uni-greifswald.de}
\affiliation{Institute of Physics, University of Greifswald, Felix-Hausdorff-Stra{\ss}e~6,\ 17489 Greifswald}

\begin{abstract}
Active matter systems characterized by the interplay of chirality and self-alignment offer a rich landscape for non-equilibrium collective behaviors and the development of autonomous materials. We present a versatile experimental platform for studying these dynamics using augmented commercial bristlebots, where custom-designed housings and elastic couplings induce a self-aligning torque and stable chiral drift. By mapping experimental trajectories to a Langevin-type model, we characterize the single-particle dynamics. In circular geometries, we show that the stability of edge currents is governed by the interaction between intrinsic particle chirality and handedness of the edge current. Furthermore, we demonstrate that transport can be geometrically rectified using a nautilus-shaped obstacle acting as a doubly chirality-sensitive ratchet. Finally, we explore the collective dynamics of rigidly linked assemblies, observing spontaneous mode-switching between translational and rotational states in active solids. Our results provide a robust framework for experimental studies in active gases and illustrate how geometric constraints can be used to program complex transport properties in active systems.\end{abstract}

\maketitle

\section{Introduction}

Coarse-graining is a necessary element in the study of complex systems in the most general sense; reduction is a prerequisite to learning in general, but a full description of every degree of freedom is neither achievable nor, crucially, desirable. Any coarse-graining procedure will result in effective dynamics, new equations of motion that are not bound by the same constraints as truly microscopic dynamics: they can be stochastic with non-conservative or even non-reciprocal forces. One class of such effective dynamics is active matter~\cite{ramaswamy2010,menzel2015,tevrugt,vrugt2025}, systems in which energy is inserted on the smallest considered scale (typically on the levels of particles or agents) which affects the dynamics by being used to generate forces or motion (self-propulsion). Such systems are manifestly outside of equilibrium as they are constantly driven on the smallest scale. A particular alluring aspect of active matter, seemingly in spite of its rather conceptual nature as a specific branch of nonequilibrium physics, is that it allows for very tangible, yet meaningful experiments~\cite{ning2024}. One big branch being realizations of self-propelled particles by way of bristlebots: small entities that have broken fore-aft symmetry which is turned into motion due to the interplay of friction and an oscillatory drive (either by an internal motor or by external vibrations of a ground plate).

{In recent years, there has been an ongoing effort to increase the tractable amount of complexity in the paradigmatical Vicsek-like or active Brownian particle models that have been the traditional workhorses~\cite{chate2020} of active matter physics.} In this letter, we present a  step in the direction of increased complexity by studying the behavior of self-aligning~\cite{baconnier_2025} chiral particles~\cite{liebchen}. Chirality in the self-propulsion, i.e., a constant drift of the particles to turn in a specific direction, is the natural extension of the Active Brownian particle model, wherein the propulsion direction is in the free case only subject to rotational diffusion. More importantly, it is reflective of the practical reality that avoiding relevant chirality requires a level of symmetry or control that is not feasible. For example, there has been recent evidence that chirality is relevant for pedestrian motion~\cite{echeverria2026}, and in the realm of artificial systems it is well-known that the asymmetric weight distribution in a commercially produced bristlebot, the Hexbug nano, induces a chirality. More often than not, however, this is ignored. Here, we instead focus on this aspect. A practical benefit is that it allows for a compactification of the experimental system as the particles' reach is somewhat bounded naturally. We are interested in the regime of an active gas: particles that explore the entire system and interact by means of collisions with finite duration. Such systems display weak correlations and are, therefore, good candidates for analytical treatment, e.g. within active kinetic theory~\cite{ihle2023,boltz2024,boltz2026}. {Self-aligning suppresses free clustering and, therefore, extends the gas regime making the present design a good choice for gas-focused studies~\cite{musacchio2025}.} The connection to actual experiments is particularly vital in active matter as the generalized structure of the dynamics a-priori allows for a greater variety in imaginable dynamics, the study of which should be focused on realizable systems. As a measure to avoid aggregation and stay in a gas-like state, we augment the bristlebots by a custom-made housing. Interestingly in the pursuit of an active gas, we find that the best design for our purposes features a self-alignment mechanism, i.e, there is a finite coupling between the direction of the active particles' actual motion which is the consequence of (steric) interactions between particles and the direction of self-propulsion. Including this level of inertia into the dynamics, endows it with a degree of memory that allows for genuinely novel behavior both on the individual and the collective level~\cite{baconnier_2025} and makes it generally more amenable to tuning. 

In this letter, we present a well-documented experimental platform with direct experimental results. In particular, we show how transport in these chiral active systems is enabled or suppressed depending on the relative chirality of particle, motion, and environment.

\section{Methods}

A fairly generic analytical model of self-propelled particles in two dimensions with positions $\vec{r}_i$ ($i=1,\ldots,N$) and internal orientations that are encoded by angles $\phi_i$ featuring steric interactions of the spatial degrees of freedom as well as the self-alignment and chiral dynamics in the angles that are the focus of this work is given by the following set of equations
\begin{subequations}
    \label{eq:model}
\begin{align} 
\dot{\vec{r}}_i =& v_0 \vec{n}(\phi_i) + \vec{F}_{i} + \sqrt{2 D_s} \vec{\xi}_i \\\propto& (\cos \psi_i, \sin \psi_i)^\mathrm{T} \\
\dot{\phi}_i =& \zeta T_i(\psi_i-\phi_i) + \omega + \sqrt{2 D_r} \eta_i \\
\langle \xi_{i,\alpha}(t)\xi_{j,\beta}(t') \rangle = & \delta_{ij}\delta_{\alpha\beta} \delta(t-t')\\
 \langle \eta_i(t)\eta_j(t')\rangle =& \delta_{ij}\delta(t-t') \text{.}
\end{align}
\end{subequations}
Herein, $v_0$ is the strength of the self-propulsion which acts in the direction $\vec{n}(\phi)=(\cos\phi,\sin\phi)^\mathrm{T}$ corresponding to the internal angle $\phi_i$. As there are also direct forces $\vec{F}_i$ acting on the particles, the direction of the actual velocity encoded by the angle $\psi_i$ does not necessarily coincide with the direction of the self-propulsion. In fact, there is a coupling between these two in form of  self-aligning torque $T_i$. For an active gas the non-self-aligning case, $\zeta=0$, would suffice; it is explicitly included here, because it will turn out to be experimentally relevant. The final elements of the dynamics are a chiral drift $\omega$ as well as spatial and orientational white noise terms.
Non-dimensionalization by rescaling times $t \to t \omega$ and positions by $\vec{r} \to \vec{r}\omega/v_0$, reveals that the angular dynamics is dominated by two dimensionless numbers $\gamma = \zeta/\omega$ and $D=D_r/\omega$. We are interested in a parameter regime where both self-alignment and chirality are relevant suggesting $\gamma \approx 1$ and dominant over the diffusive noise, i.e., $D\ll 1$. In general, this rotational noise will be the relevant source of noise as the spatial diffusion is only relevant on a lengthscale $\ell_s \sim \sqrt{D_s/D_r}$ and the actual value of $D_s$ will be very small in any realization of dry active matter~\cite{chate2020} that is dominated by friction. Interestingly, this friction is of static origin for bristlebots~\cite{antonov}.

In the aim of implementing a chiral active gas experimentally~\cite{wagnerthesis}, we opted to use the already mentioned commercially produced bristlebots of type Hexbug nano as active element. For a table-top experiment, they are however not ideal. For one, they tend to get stuck at walls and accumulate there and, for two, the chirality is fairly weak and, thus, the length scale on which it is observable $\ell_\omega \sim v_0/\omega$ is large compared to the size of the experiment. We use  a circular arena with $R=25$cm as base design of our experiments due to limitations with the camera field-of-view, table space and 3D-printing capabilities. 

\begin{figure}
\includegraphics{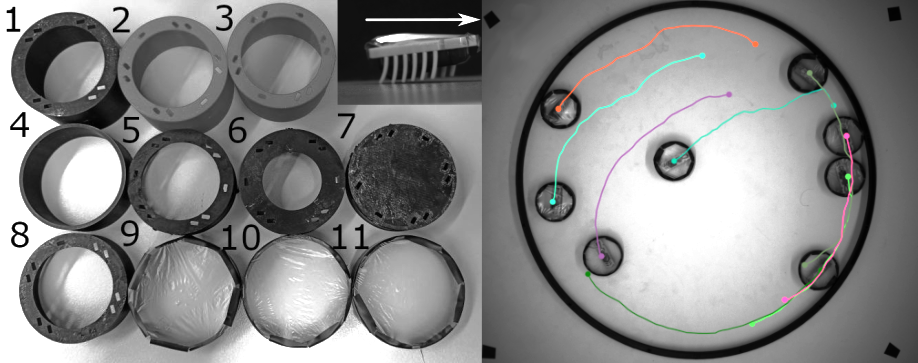}
\caption{Left: Candidate designs for the housing in which a Hexbug nano (inset) is placed. The designs differ in wall-height and design of the lid as well as the lower rim. A full evaluation of these designs is presented in the supplemental material~\cite{supplmaterial}.  Right: Experimental setup in a simple circular arena with tracked trajectories overlaid (colored lines). The photographs were changed to gray-scale and slightly brightened to increase printability.}
\label{fig:1}
\end{figure}

A direct approach that addresses both of these issues is to encapsulate the bristlebots with circular housings, effectively creating new active particles with circular symmetry (which in the future might alleviate analytical calculations). The rings are produced by a 3D printer (using Tough PLA from UltiMaker). The printing costs for a single
ring are around 1 Euro. Our design goal at this stage is to create active particles that move chirally, but still are fast, so that activity is still a relevant factor and the observable dynamics allow for relevant insight. 
The bristlebots at the center of our particles operate via battery power. This limits observation time, so as not to see aging effects from draining power. Before all measurements, we ensure that all particles are properly charged. Another design difficulty is that the bristlebots have an individual power switch on their body, this means that there is somewhat reduced control over initial conditions. We present a survey of housing designs that have been explored in fig.~\ref{fig:1} (left). The radius ($\ell_h= 2.7$cm with a wall thickness of $0.2$cm) is designed such that it is as small as possible while still allowing the bristlebot to change directions with the housing. This means that the augmented bristlebots as a unit is able to turn on the spot and, thus, the tendency to form jams is vastly reduced. In addition to varying cylinder heights, we tried various choices for the lower (ground) border and lid. Generally, any lid will transfer the bristlebots jumping motion to lifting the ring as well and, thus, reduce the effect of static friction with the ground. In agreement with prior studies~\cite{chen2023,wang2023}, we find that a hard lid induces a rather large persistence time while mostly suppressing the chirality, leading to particles that move mostly straight for long distances on the order of a meter. We find the system to be in a substantially more interesting parameter space using a rather soft elastic lid. Specifically, we use commercially available high-density polyethylene foil (reported thickness of $5\,\mu$m, sold as trash bags). We infer from the video material of ref.~\cite{baconnier2022} that a similar design was used in that work. An equivalent, but slightly different design choice that has recently been employed is a softer housing that is directly friction-coupled to the bristlebot and provides self-alignment this way~\cite{carillo2025}. We show empirical velocity distributions as inferred from the displacements across subsequent frames for all designs in the supplemental material (SM)~\cite{supplmaterial}, we finally opted for design 11.

Observation was done by means of automatically processed video footage. Images are taken by a camera in gray-scale at a resolution of $2464\times 2056$ (partially reduced by undersampling to reduce memory consumption). Most measurements span  around $2000$ frames taken at a frame-rate of $25$Hz.  As a gauge to convert between pixels and cm, an image of a
reference with known length is taken. The identification of particles within a single frame is achieved by means of a circular Hough transform\footnote{Specifically, we make use of the \textsc{imfindcircles} routine of octave's~\cite{octave} image package.}. This identifies the location of particles with their centers and does not reveal any information about the orientation of the internal bristlebot. However, it is a very robust method and, for example, reflections of light on the covering foil do not lead to misidentifications or missing frames. Another positive consequence of this approach is that all identified positions are fairly far apart from each other as the housings will not overlap and tracking of particles over multiple frames is straightforward. We can safely identify a particle in a frame with the closest particle in the previous frame, as the expected displacement in between frames at a frame rate of $25$Hz is below 1cm and thus substantially smaller than the diameter of the particle housings. At the moment, our imaging capabilities did not suffice to accurately infer the position and alignment of the internal bristlebots (and hence the direction of the self-propulsion) automatically. In future work, exploring this internal structural flexibility generated by the composite nature of our particle design~\cite{ketzetzi2025} will be an interesting direction. On a conceptual level, we expect this structural softness will lead to effectively aligning interactions in collisions~\cite{grossman2008,menzel2012,grossmann2020,das2024}.

\section{Results} In the following, we present results for the free (within a circular domain) dynamics as well as within more structured environments that highlight the interplay between intrinsic geometry of the internal dynamics and extrinsic geometry of the environment. 
\subsection{Free Dynamics}

\begin{figure}
\includegraphics[width=\linewidth]{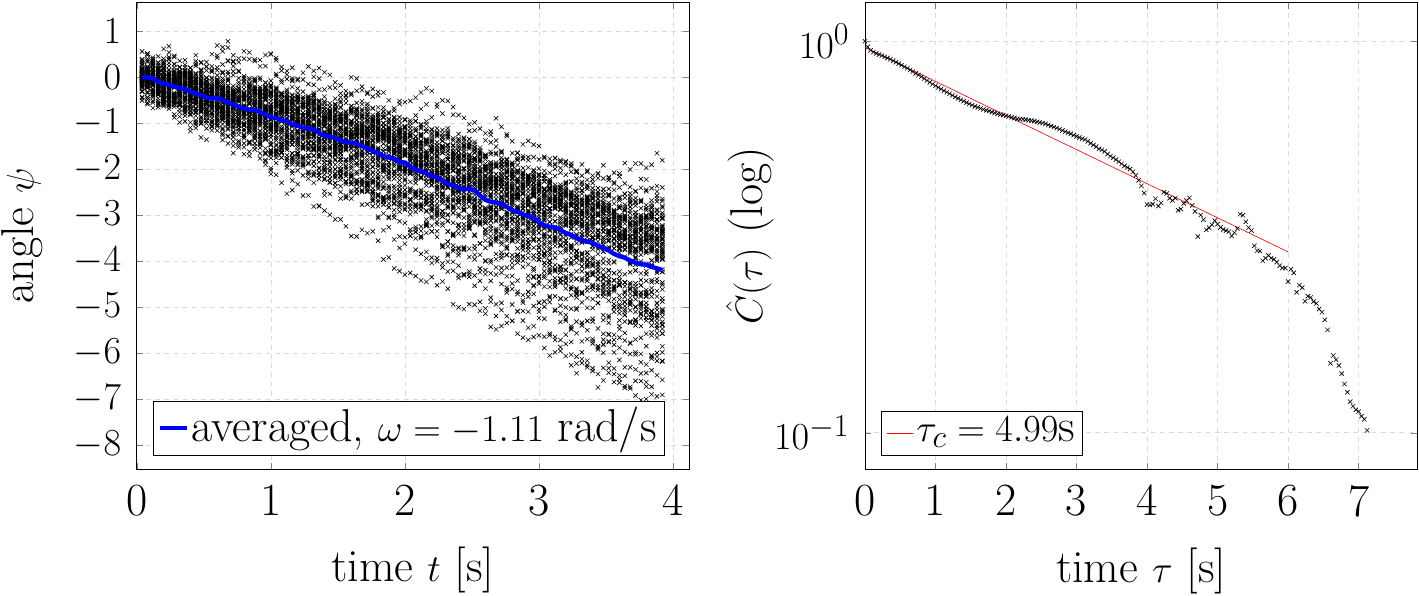}
\caption{Left: Scatter-plot of experimentally determined angle differences as a function of time. In line with the data processing to free dynamics, we identify the direction of motion with the direction of self-propulsion here. The angles are not periodically wrapped, but constructed from the increments at each time step. On the ensemble level (solid blue line) an overall line, consistent with a constant chirality emerges.  Right: Adjusted velocity auto-correlation function (see main text). We see that we can clearly identify the effect of rotational noise (red line). }
\label{fig:2}
\end{figure}

From the empirical velocity distribution, we infer that our particles have a typical speed of $v_0 \approx 15.75$cm/s. This measurement is the only aspect of the dynamics, we can infer from single-time measurements. The other aspects, chirality and noise, require analysis of the correlations between multiple times. In quantifying the decorrelation of the velocities we have to disentangle three effects that change the velocity orientations over time: interactions with the boundary or other particles, the chirality, the noise. We can account for the first by a selection of trajectories, but the latter two require slightly more attention. Without noise and interactions, a particle that follows eq.~\eqref{eq:model} with a starting angle of $\phi$ would have an angle of $\phi(\tau)=\phi+\omega \tau$ after time $\tau$ (discarding periodic wrapping for now, to keep the notation simpler). Thus, a sensible measure of noise induced decorrelation is the {\em adjusted velocity autocorrelation function}
\begin{align}
C(\tau) &=  \langle (\mathsf{M}(\omega \tau) \vec{v}(t))\cdot \vec{v}(t+\tau)\rangle_{t} -\langle \vec{v}(t)\cdot \vec{v}(t)\rangle_{t} \label{eq:cadjust}
\end{align}
wherein $\langle \cdot \rangle_{t}$ indicates an averaging over both the starting time $t$ and the experimental ensemble and $$
\mathsf{M}(\theta)=\begin{bmatrix}\cos \theta &-\sin \theta \\\sin \theta &\cos \theta \end{bmatrix}
$$ is the matrix which conveys a rotation by an angle $\theta$. In fig.~\ref{fig:2}, we show that this approach does indeed allow us to infer a meaningful quantification of the rotational noise in terms of a single relaxational time constant $\tau_C$, such that $C(\tau) \sim \exp{(-\tau/\tau_C)}$. To this end, we identified a single drift constant $\omega$ from the linear trend in (non-periodically wrapped) time-series of the angle $\phi$ (left panel). We found $\omega = - 1.11$rad/s (sign omitted in subsequent scale considerations). This implies a typical length scale on which turning occurs of $\ell_\omega=v_0/\omega \approx14.2$cm which is of the same order as the system size $R$ and, thus, we expect that the system boundaries are important, while the entire system is explored. This is consistent with our observations. Using this value for $\omega$ as well as eq.~\eqref{eq:cadjust}, we find $\tau_C \approx 4.99$s. This corresponds to a decorrelation over a traversed angle of $\omega \tau_c\approx 1.8 \pi$ and length of $\ell_C \approx 79$cm, which is substantially larger than our system size indicating that we are indeed characterizing epicyclical, self-intersecting trajectories. 

On the level of treating the augmented bristlebots as a unit, there will be an alignment component to steric interactions with other particles and system boundaries. Inferring this interaction from experiment is difficult, as we have limited access  to (or initial control of) the self-propulsion direction. We have implemented an in-silico reproduction of our system (see SM~\cite{supplmaterial})  that circumvents this inference by keeping the external housing and internal bristlebots as individual entities that interact via Hertzian contact mechanics. The specific shape of the friction-induced self-aligning torque is equally rather inaccessible. However, there is a fairly general symmetry-based argument~\cite{baconnier_2025} leading to a dominant dependence like $T_i \sim \sin(\psi_i-\phi_i)$. The strength $\zeta$ of the self-alignment can be inferred from a simple experiment by imposing a velocity by hand\footnote{see Video in the SM~\cite{supplmaterial}} and observing the time it takes the bristlebots orientation to  reorient to this direction. In agreement with ref.~\cite{baconnier2022}, we find that $\zeta\approx 5/s$ is reasonable. This and its connection to textbook nonequilibrium statistical mechanics problems (the washboard potential of Josephson junction~\cite{Ambegaokar,stratonovich} fame) is also discussed in the SM~\cite{supplmaterial}.

The observed mean square displacement (MSD) follows a power law, $\langle \Delta \vec r ^2\rangle \sim t^{2a}$ with $2a \approx 1.68$ (see SM~\cite{supplmaterial}). The deviation from the purely ballistic limit  is explained by the chirality. In the SM~\cite{supplmaterial}, we use standard stochastic calculus to directly compute the velocity correlations from a free version of eq.~\ref{eq:model} and the MSD~\cite{vanteeffelen,Pattanayak_2024,Littek_2026,kiechl2026} from it. We show that a non-power-law MSD emerges that can resemble an effective power law in the correct region. On large times (corresponding to inaccessible large displacements), we expect a purely diffusive regime for symmetry reasons. 
 
\begin{figure} 
\includegraphics[width=\linewidth]{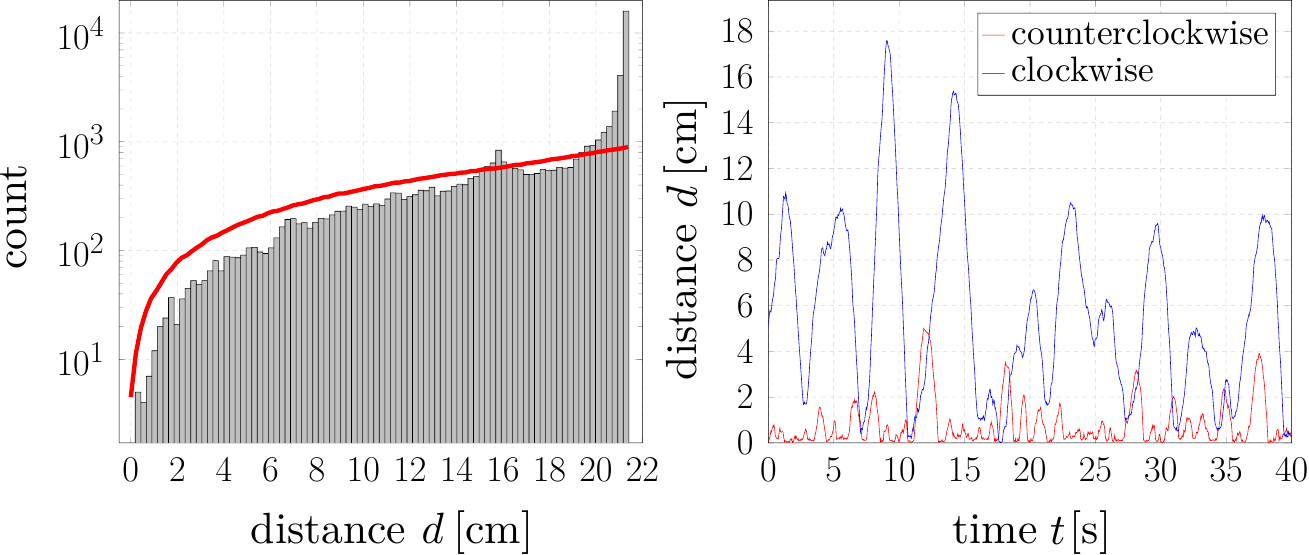}
\caption{Left: Radial distribution of bristlebots, showing a significant accumulation at the system walls. We overlay (red line) synthetic data (Monte Carlo simulation of non-overlapping circles placed randomly) as a comparison highlighting the large excess close to the wall. Direct wall contacts as well as the first shell are clearly visible. {We study the wall interaction dynamically by means of wall residence times in the SM~\cite{supplmaterial}.}  Right: Timeline of the wall distance for a single clockwise and a counterclockwise particle.}
\label{fig:3}
\end{figure}

There are, in general, quenched inter-bot differences. For example, if we constrain the motion to a single circular file (akin to refs.~\cite{tian2018,patterson2023,knippenberg}), the overall motion will always be set by the same slowest bot. These differences are generally, however, not as apparent, and we can treat the augmented bristlebots as a homogeneous ensemble. {We limited the intra-ensemble variability by choosing the $N\leq 8$ bots in actual use from a larger ensemble of $N_{\text{all}}=25$ that are close to each other in dynamics, see also SM~\cite{supplmaterial}.}

Overall, we find that our experimental platform does indeed reproduce the dynamics of eq.~\eqref{eq:model} with $v_0 \approx 15.75$cm/s, $\zeta \approx 5$/s, $\omega \approx 1.11$rad/s, $D_r \approx 0.2/$s and a particle size (setting a length-scale for steric interaction) $\ell_h \approx 2.7$cm. The relevant dimensionless quantities therefore are $\gamma \approx 4.5$, $D \approx 0.18$. The fact that $\gamma>1$ is physically relevant as it means that the alignment under persistent forcing in a given direction will dominate over the chirality. We are using an arena with $R=25$cm and up to 8 particles, this corresponds to an occupied area fraction of $\phi_A \approx 0.1$. Importantly, we are able to infer all model details but the (very fast) steric interactions by direct observation and our measurements, therefore, also allow for future in-silico experiments using actual physical values. As noted before, the velocity, chirality and self-alignment strength are subject to design details (of the housing and the elastic lid) and can thus be tuned. In principle, it would also be possible to almost completely eliminate the chirality on the bristlebot level, either by choosing a different bot altogether or by adding a small counterweight that counteracts the structural imbalance~\cite{balda2024}.

\subsection{Circular arena} We use a 3D printed circular arena with upright walls. In fig.~\ref{fig:3}, we show that this wall does relevantly affect the behavior, with a peak in the radial distribution at wall contact (and a minor peak two radii from the wall). This dominance of system boundaries is for one to be expected given that the intrinsic dynamics set a length scale $v_0/\omega \approx 14.2$cm of a comparable lengthscale as the system dimensions, but also more generally fairly typical in active matters~\cite{deblais2018,leoni2020}. However, we also see that the design choices are working to their goal, there is no jamming, the entire system is explored and there is a circular flow around the exterior wall.  The tendency to be close to the walls leads to a density structure similar  (see SM~\cite{supplmaterial}) to motility induced phase separation observed in other active systems with hard-body interactions. However, this system is always in a predominantly moving state and the small system sizes studied here limit the notion of densities a lot. Chirality endows active matter~\cite{liebchen,mecke} with new possibilities for emergent behavior.  One  emergent phenomenon in chiral fluids are edge currents: small scale vortex-like flows coagulate to large scale flux at the boundary of a system. These have previously been realized in bristlebot-based experiments~\cite{barois2020,yang2020,caprini2025}. In our case, the rotational motion along the border is mostly a consequence of the persistence of the motion rather than the chirality. Even a non-chiral version of our system would mostly move along the walls~\cite{deblais2018,leoni2020}, but the chirality alters the preference and stability of these motions, see fig.~\ref{fig:3} right.   This is a consequence of a symmetry breaking, there is an explicit difference between the wall being to the left or to the right of the particle (with respect to its direction of motion). We can observe states of collective motion with both handedness. If the handedness of the transport and the particle disagree (red-line, ccw motion) the particles constantly turn into the wall: the edge current is stable and particles stick closely to the wall; in the other case, turning away from the wall leads to large fluctuations in the wall distance. {We can observe the same distinction dynamically by observation of wall residence times, see SM~\cite{supplmaterial}.} It is a direct consequence of the self-aligning design, which leads to effective tangential wall torques from softer guiding interactions instead of the bristlebots directly hitting the wall and potentially getting stuck,  that the edge currents are fairly directly observable in this system.

\subsection{Chiral environments}
\begin{figure}
\includegraphics[width=\linewidth]{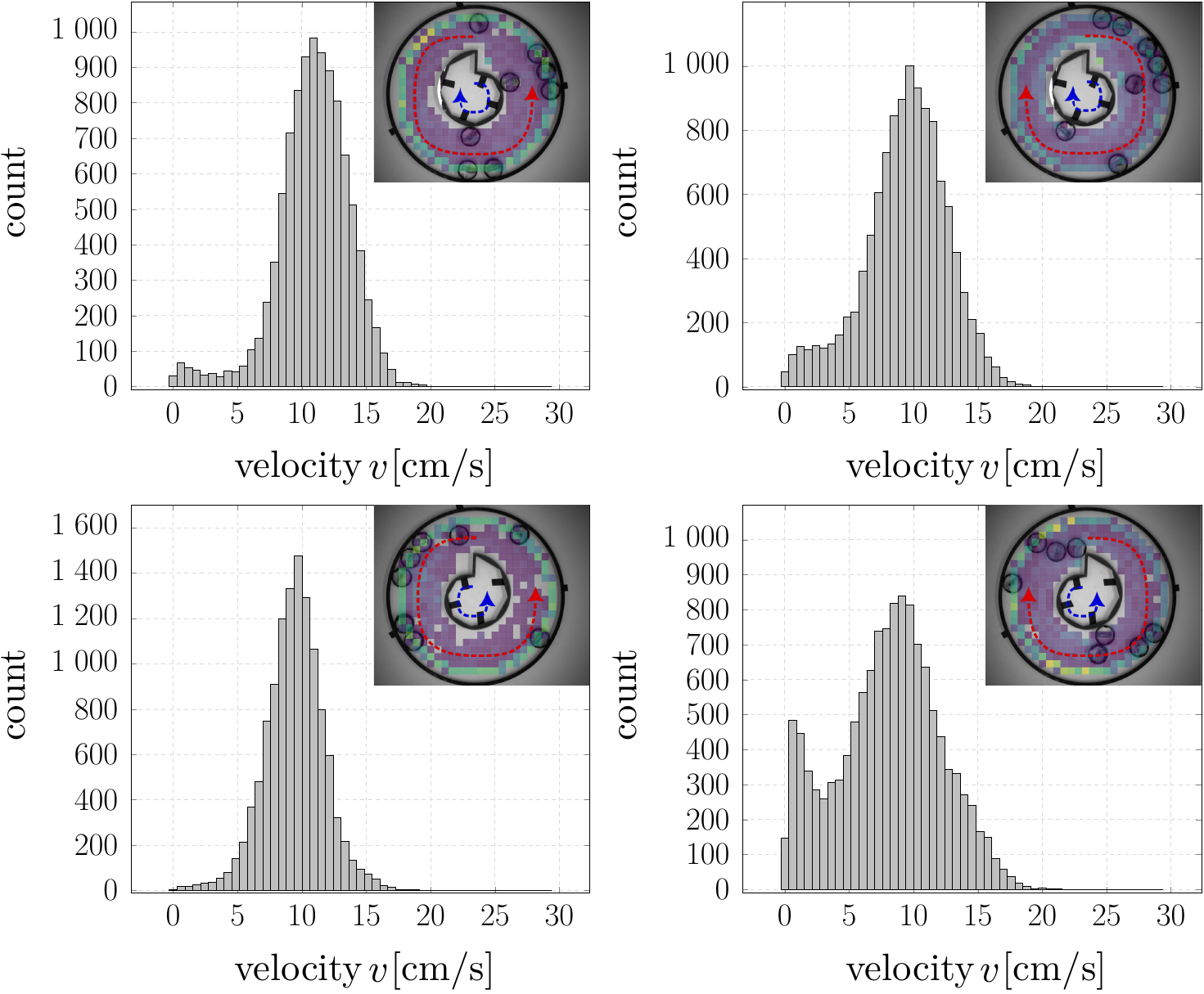}
\caption{Interaction of  environmental chirality with the intrinsic chirality of the active particles and the chirality of the current state of motion. We place a chiral object, in form of nautilus that can be parametrized by a radius that is a linear function of the angle, in the center of the circular arena. We indicate the handedness this implies by a blue arrow.  Collective motion of active particles in a circular configuration of defined handedness (red arrow) is induced. If both chiralities agree with the intrinsically preferred direction (lower left panel), there is global transport with little jams as indicated by an empirical velocity distribution that vanishes at small velocities. In the other cases, particularly in the case of clockwise motion in a counterclockwise oriented environment (lower right panel), the onset of jamming is clearly visible as a pronounced excess of near-zero velocities. {In addition to the handednesses, the insets also show a heatmap visualization of the coarse-grained empirical probability function, see SM~\cite{supplmaterial} for details.}}
\label{fig:4}
\end{figure}

Inducing specific dynamics of chiral active matter by means of a chiral environment has been previously explored for bristlebots~\cite{barois2020,xu2022,li2022,balda2024,chan2024}, we add to this by constructing a bottleneck~\cite{patterson2017,barois2020} to the dynamics that is sensitive to the particle chirality. We place a nautilus-shaped (or, precisely, an Archimedean spiral) obstacle in the center of our arena, see fig.~\ref{fig:4}. The shape is easily parametrized in polar coordinates by a radius $r(\phi)=r_0+a\phi$ that has a linear dependence on the angle. The sign of $a$ differentiates the two possible handednesses, which in the experiment corresponds to flipping the obstacle on its head. The effect of the nautilus is that it constricts the flow around the center of the arena to a rather narrow (approximately the width of two particles) bottleneck at one point (the top in the insets). The approach to this bottleneck is vastly dependent on the direction of motion of the particle: from one side there is a smooth narrowing whereas the other side has a rather large open space. This design leads to a relevance of the fluctuations from the direction of motion, the space on the open side is sufficient for the particle to turn around, which can induce a blockage of particles. This leads to the four cases discussed in fig.~\ref{fig:4}, where we consider rotational motion of both orientations in both chiralities of the environment. We show empirical velocity distributions to highlight the fundamental difference in transport due to the obstacle in these four scenarios. Geometrical rectification~\cite{nikola2016,reichhardts,lee2021}, enhanced transport by means of an asymmetry in the environment is a genuinely non-equilibrium effect as it violates detailed balance. The chirality ratchet effect at play here is a direct consequence of the chirality-induced differences in volatility of the edge currents that we show in fig.~\ref{fig:3} (right): The most pronounced presence of small velocities, indicative of jamming, is observed in the case of clockwise (unstable) moving ensemble with a counterclockwise oriented obstacle (lower right). In this case the fluctuations entropically increase the effect of the bottleneck. Flipping the obstacle (upper right), removes most but not all of the jamming. For counterclockwise (stable) moving ensembles, we again find that transport is enhanced by matching the obstacle and motion handedness. The bottleneck is fairly irrelevant in the best case (lower left).

Overall, we show that self-aligning particles respond to external geometry smoothly and this allows for a great level of passive external biasing.  We show transport regulation by means of an active ratchet effect. This effect is particularly pronounced if the fluctuations due to the intrinsic chirality are important. Chirality can be suppressed in some applications, but is generally ubiquitous~\cite{liebchen,mecke} and, thus, the handedness of interactions should be considered an important aspect in future endeavors.

\subsection{Rigid constraints}

\begin{figure}
\includegraphics[width=\linewidth]{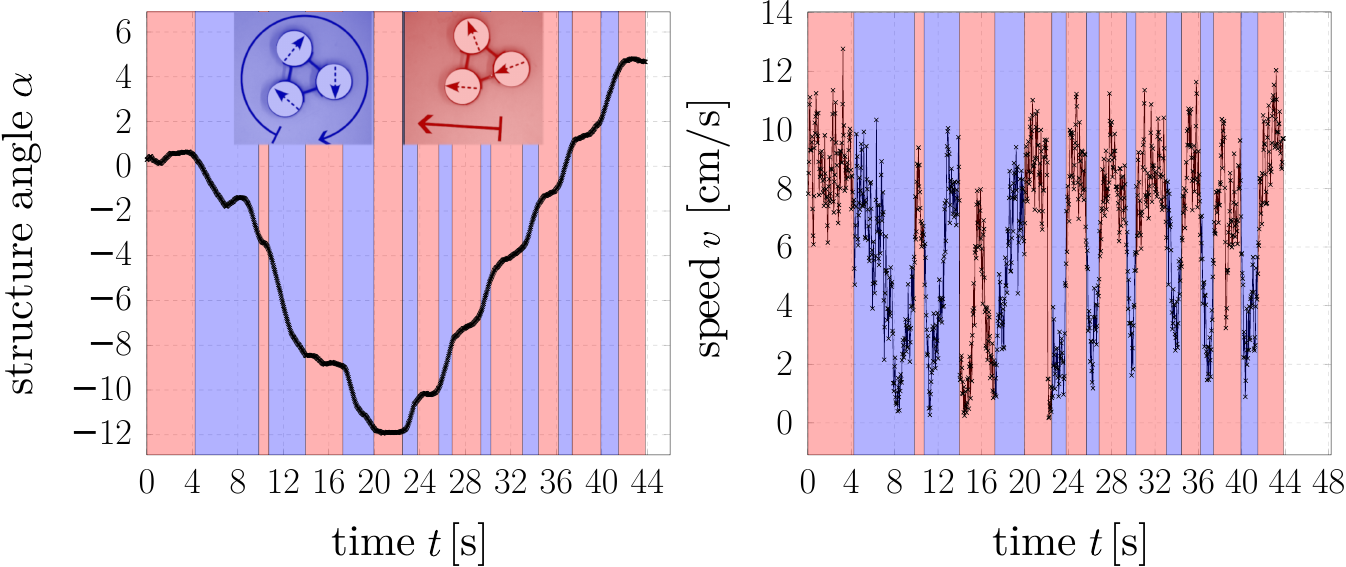}
\caption{Left: Time-series of the orientational angle of the triangular structure. This is taken to be the angle of the displacement vector between the center of mass and a fixed particle to a spatially fixed axis. The dynamics shows a clear distinction into two states: a rotating state (left inset) in which the group angle changes quickly, and the collective motion is suppressed and a moving state (right inset) in which the group angle is mostly constant, and the collective velocity is large. Right: Time series of the corresponding center-of-mass velocity of the triangular structure, showing the same bipartite structure that is evident in the orientational angle. }
\label{fig:5}
\end{figure}

In the previous section, we {biased} the motion by an explicit bottleneck in space. However, one can also design bottlenecks in phase-space that still allow every point in space to be visited. A particular type of such an environment is adding permanent elastic interactions to the particles. These add additional structure to the phase-space that hinders (or blocks) access to some regions, similarly to how the housing restricts the accessible phase-space regions to configurations with a fixed minimal distance between particles. Assemblies of active particles with elastic constraints are interesting for several reasons: In the stiff limit, they impose kinetic constraints that limit the motion to the zero modes of the configuration. In ref. \citenum{hernandezlopez2024}, Hernandez-Lopez et al. followed this idea to analytically identify that such active clusters (or solids) explore an effective energy landscape with minima that correspond to the collectively admissible states. We demonstrate this, by considering three particles linked in an equilateral triangle. This was implemented by means of 3D-printed linkages between the housings. There are two zero-modes in this system: translation and rotation of the triangle.  In fig.~\ref{fig:5}, we show that we can experimentally realize this concept. The time series of the rotation of the triangle (left), shows that there are periods of almost stationary orientation (corresponding to translation) that alternate with fast changes (rotation). This is complemented by considering the overall speed of the construct (right). The stretches of nearly constant orientation correspond to large speed, whereas the center-of-mass is almost vanishing in the rotational state. Interestingly, we find little direct evidence that the intrinsic chirality of the bristlebots is affecting the collective motion, which does display some persistence across subsequent rotational phase, but no overall tendency. We connect this to the fact that the radius of the cyclical motion imposed by the triangle (fixed separation $d\approx \ell_h \approx 2.7$cm) is vastly differing to the intrinsically preferred motion even in the case of aligning handedness and the influence of wall interactions. Thus, we are concluding that this particular design offers a topological implementation of an active run-and-tumble particle. {Interestingly, the data shows little sign of the free handedness. This, however, has to be attributed to the role of wall interactions. Free simulations (see model in the SM~\cite{supplmaterial}) show clear chirality}. In future work, it will be interesting to explore to what extent the parameters of this effective motion can be {selected} by  specific design choices such as the lattice constant of the triangle or the amount of friction imposed by elastic cover. In other works, assemblies of few bristlebot-driven active particles by flexible~\cite{zheng2023} or semiflexible (buckling)~\cite{xi2024} elastic connections have been considered. These directions also allow for a wider variety in the design of the effective energy landscape explored by the active particles. {We note that a similar reduction to two collective modes can be observed in active crystals of self-aligning chiral particles~\cite{musacchio2026}.}

\section{Discussion \& Conclusion} In this work, we experimentally explored how transport within an active gas can be induced by means of bottlenecks, restrictions to the flow of particles either in actual space or in phase-space. To this end, we designed an experimental platform based on commercially produced bristlebots that are placed into custom-made housings to allow for pronounced chirality and self-aligning behavior which enhances {the opportunity for passive biasing of the overall transport}. Our work demonstrates that even minimal active units, when augmented with self-alignment and chirality, can exhibit rich, {selectable} behaviors indicating a way for designing adaptive active materials. Our observation of a chirality ratchet using a nautilus-shaped obstacle highlights the potential for passive transport design in active systems. We have shown that the rectification of particle flow is not merely a product of geometric asymmetry, but emerges from a delicate matching between the handedness of the environment and the intrinsic stability of the particles' edge currents. This sensitivity suggests that environments can be tailored to sort or filter active agents based on their individual chiral signatures, a principle that may find applications in microfluidic sorting or the design of autonomous robotic swarms. The transition between translational and rotational states in our triangular assemblies constitutes a physical realization of topological mode-switching and direct implementation of analytical ideas~\cite{hernandezlopez2024}. The fact that these 'active solids' can explore a complex phase-space, alternating between fast center-of-mass motion and stationary rotation, indicates the malleability of active materials~\cite{menzel2012}  opening the door to designing multi-component active structures where the global motility is dictated by the internal geometric zero-modes of the assembly. While we have focused on a gas-like regime, scaling these systems to higher densities may reveal novel phases where chirality and self-alignment compete to drive large-scale pattern formation or active turbulence. Ultimately, our results underscore the importance of considering the interplay between the geometries of active dynamics and (phase-)space as control parameter in the design of adaptive active matter. 
 
\begin{acknowledgments}
We thank C. Hern\'andez-L\'opez for bringing ref.~\citenum{hernandezlopez2024} to our attention in a talk at the University of Greifswald.

{\em Data availability statement:} All relevant data (inferred trajectories as well as design CAD files) are openly available~\cite{zenodo}. 
\end{acknowledgments} 

\bibliography{references_arxiv.bib}

\end{document}


\title{Supplemental Material for ``Geometry-Induced Transport of Self-Aligning Chiral Bristlebots''}

\author{Timo Wagner}
\author{Michael Himpel}
\author{Thomas Ihle}
\author{Horst-Holger Boltz}
\email{horst-holger.boltz@uni-greifswald.de}
\affiliation{Institute of Physics, University of Greifswald, Felix-Hausdorff-Stra{\ss}e~6,\ 17489 Greifswald}
\begin{abstract}Additional material in support of the main text. We present the analyzed design options as well as some analytical calculations in the reduced Brownian circle swimmer model. Finally, we show some evidence for phase-separation-like phenomenology and describe the provided video material.\end{abstract}
\maketitle

\section{Housing designs}

\begin{sidewaysfigure}
\centering
\includegraphics[width=0.24\linewidth]{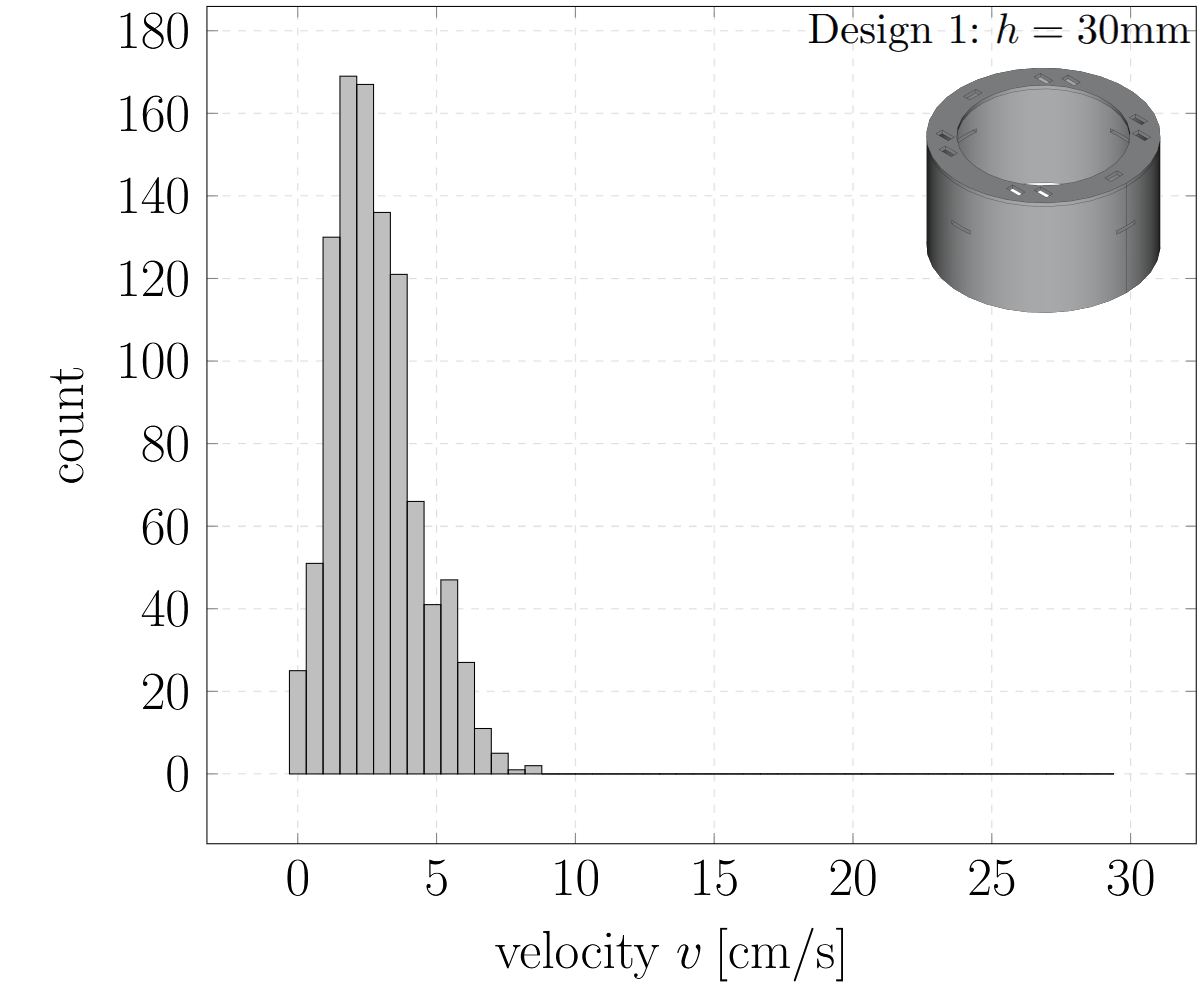}\hfill\includegraphics[width=0.24\linewidth]{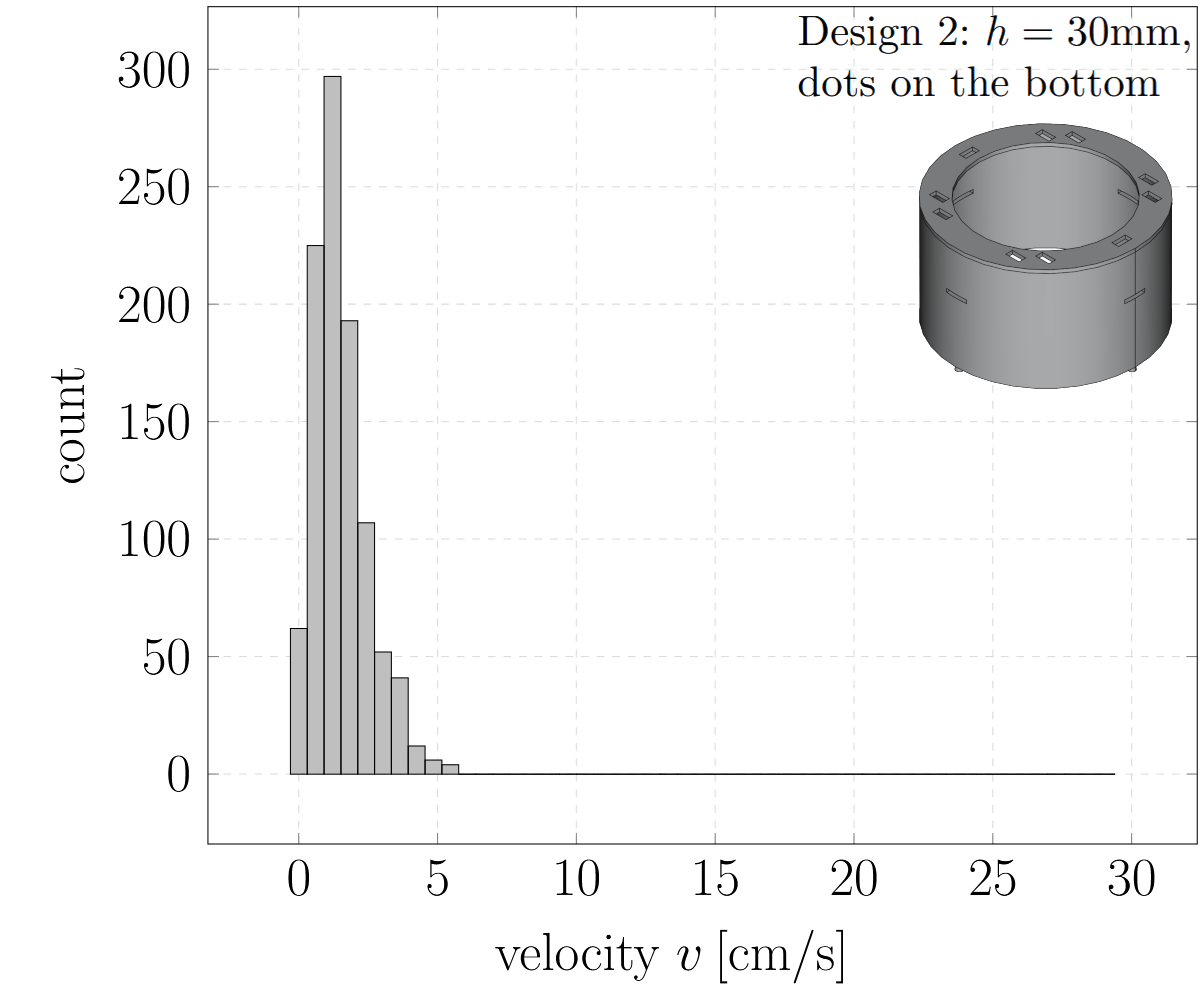}\hfill
\includegraphics[width=0.24\linewidth]{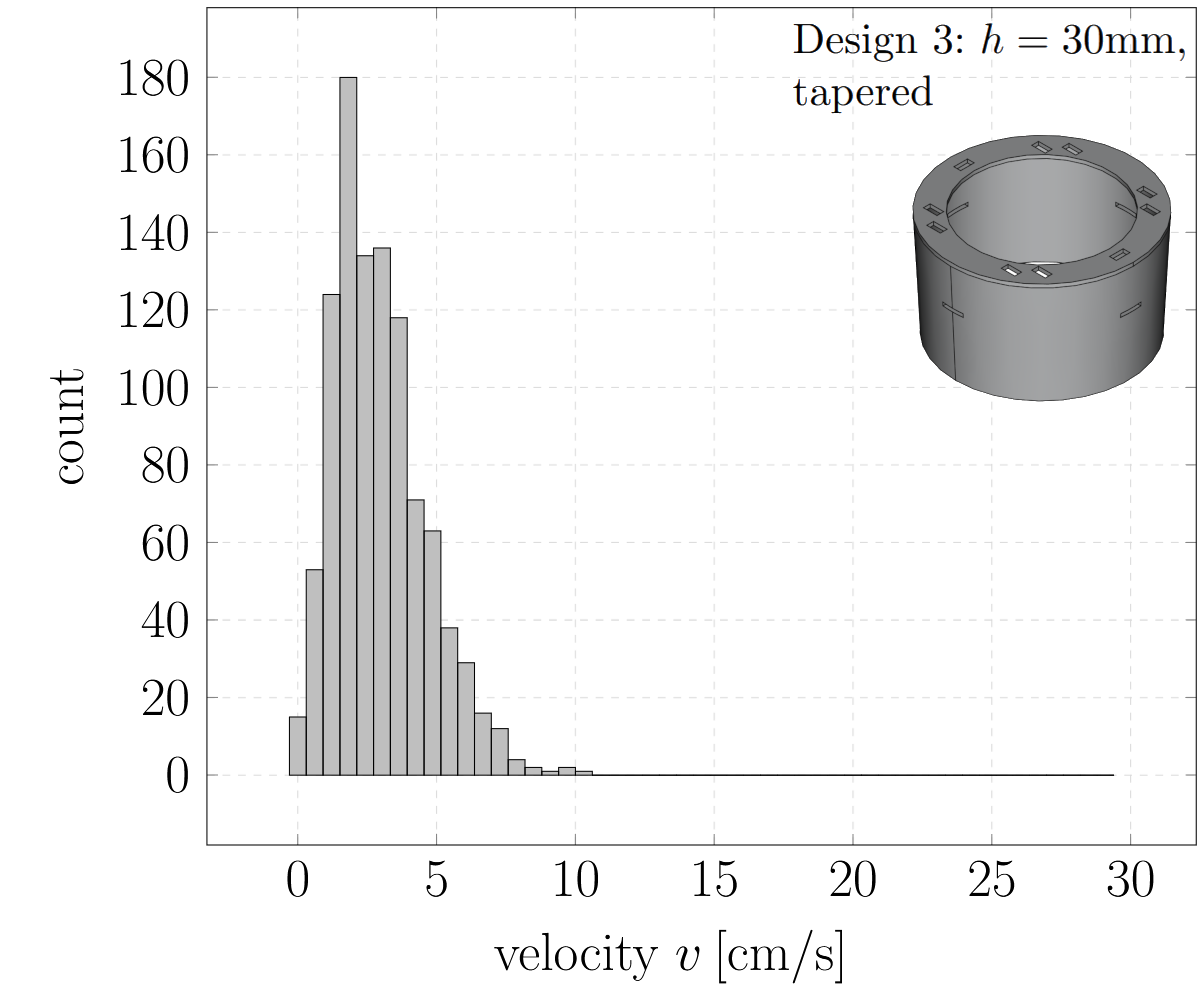}\hfill\includegraphics[width=0.24\linewidth]{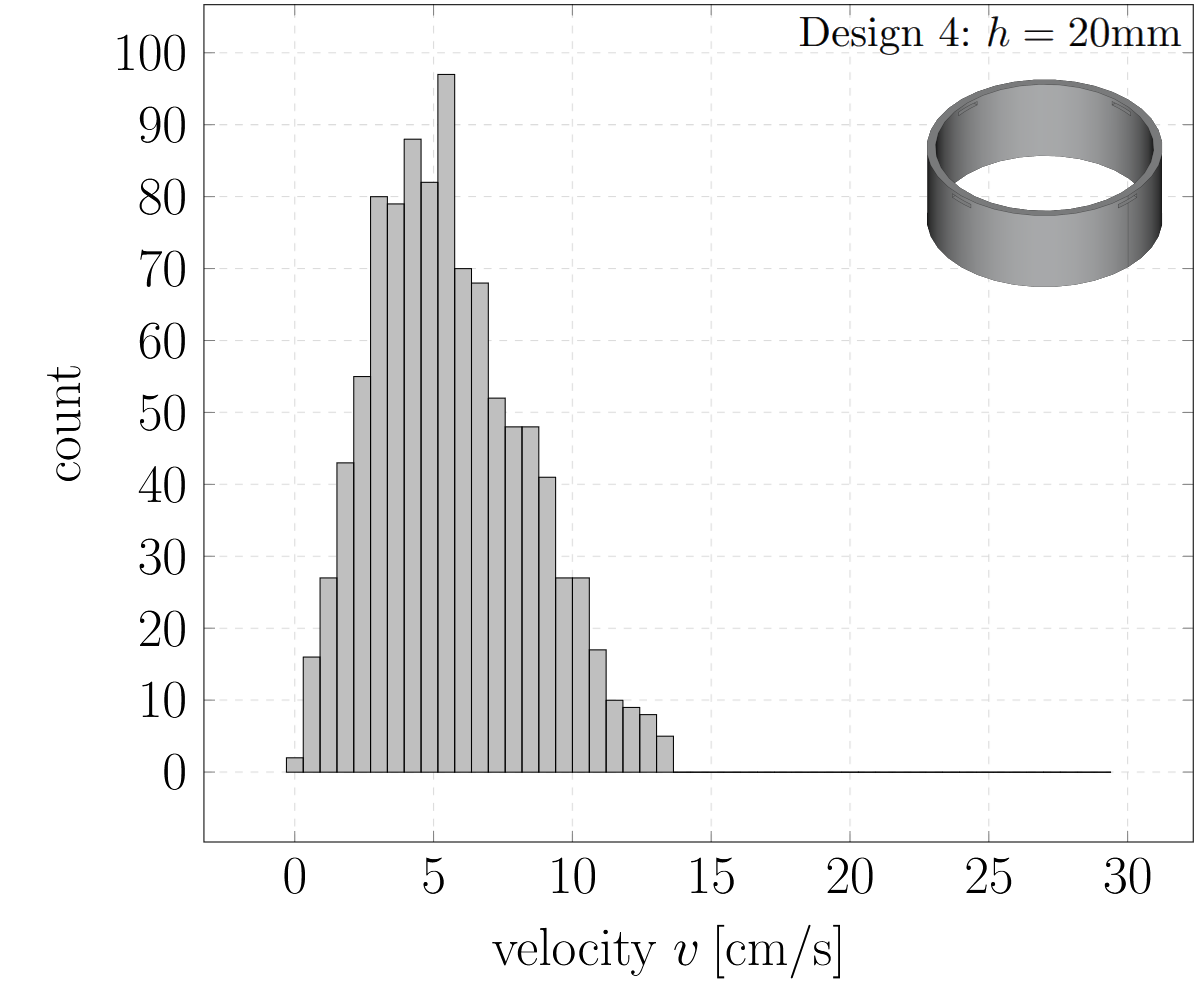}\\
\includegraphics[width=0.24\linewidth]{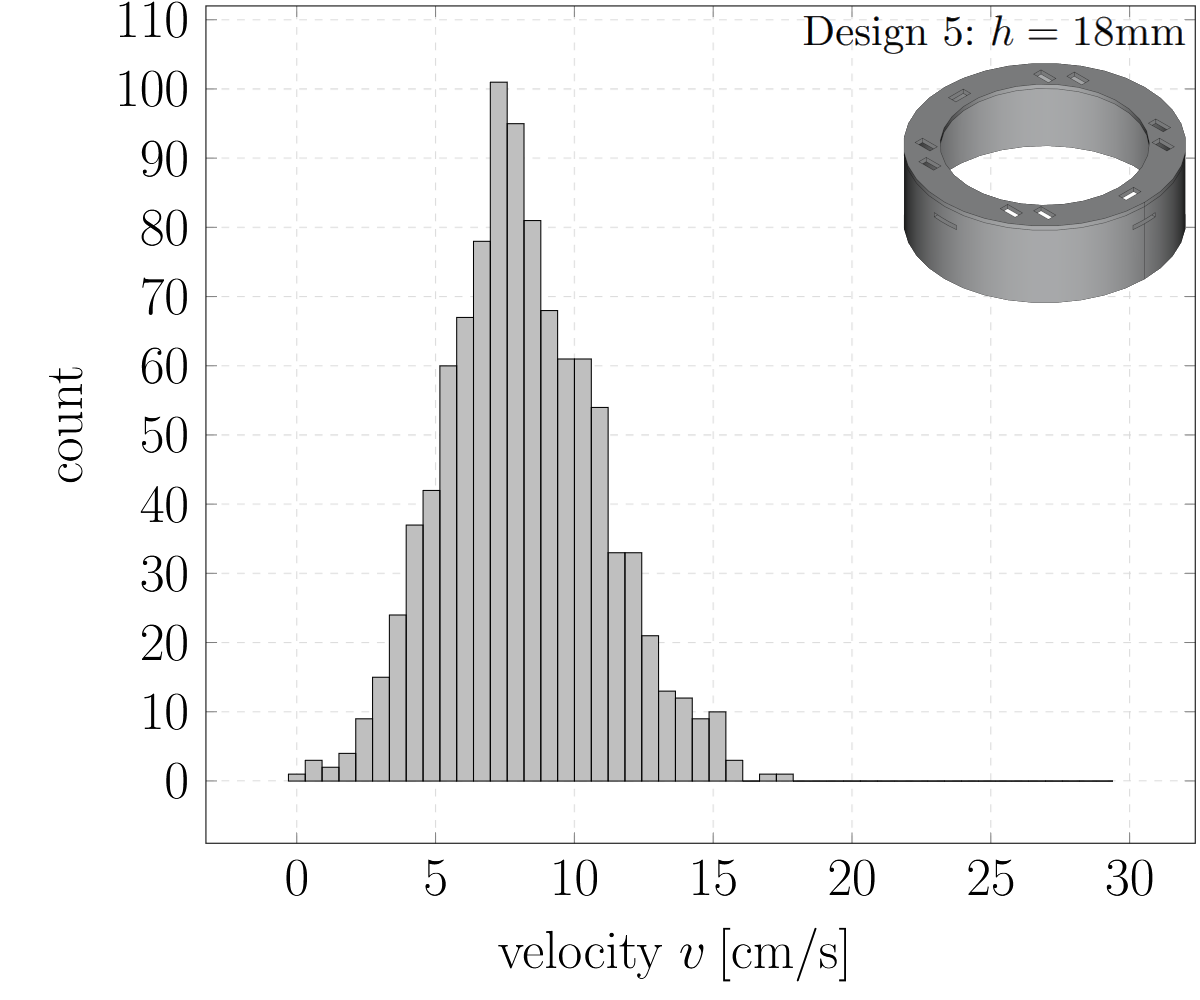}\hfill\includegraphics[width=0.24\linewidth]{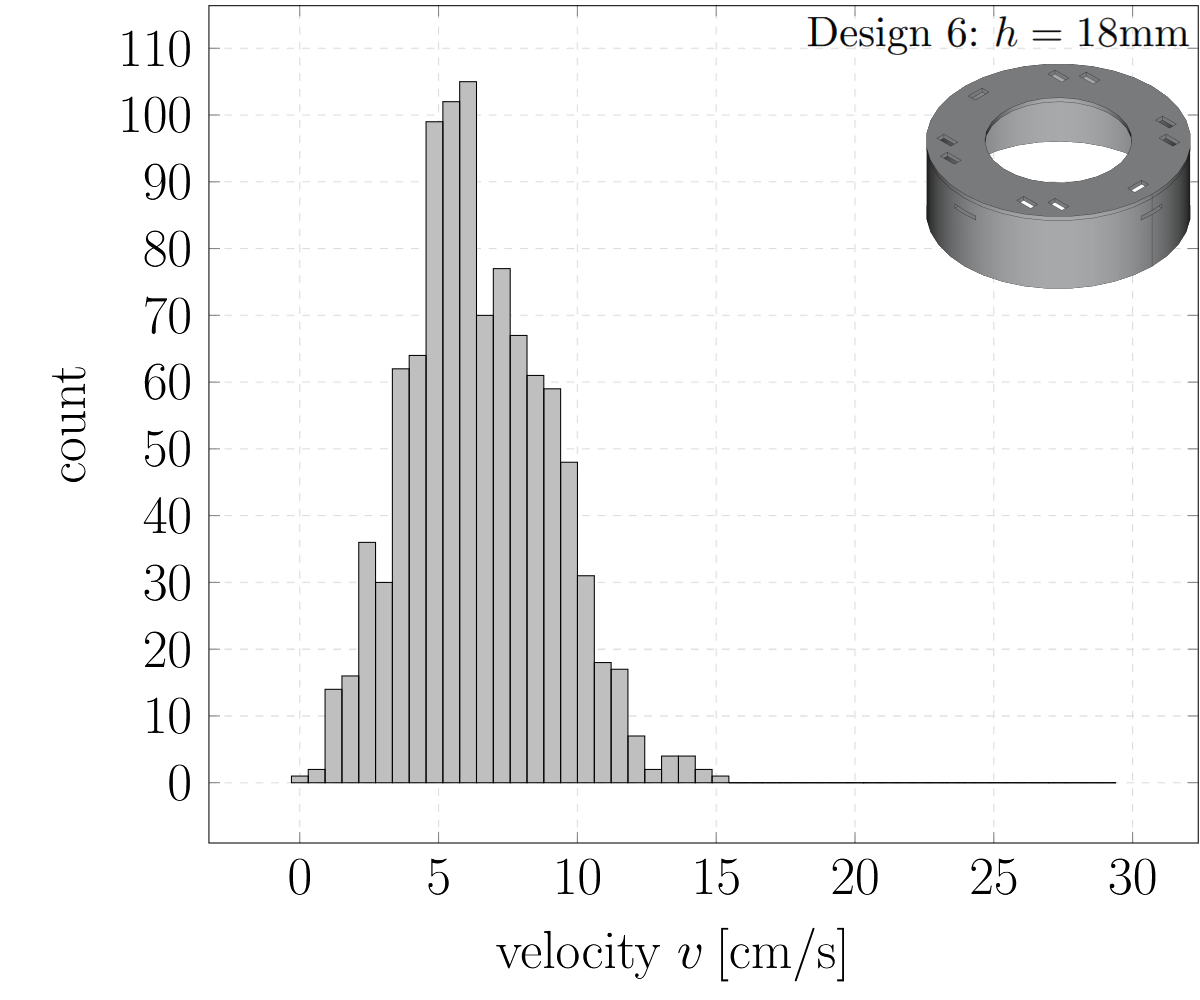}\hfill
\includegraphics[width=0.24\linewidth]{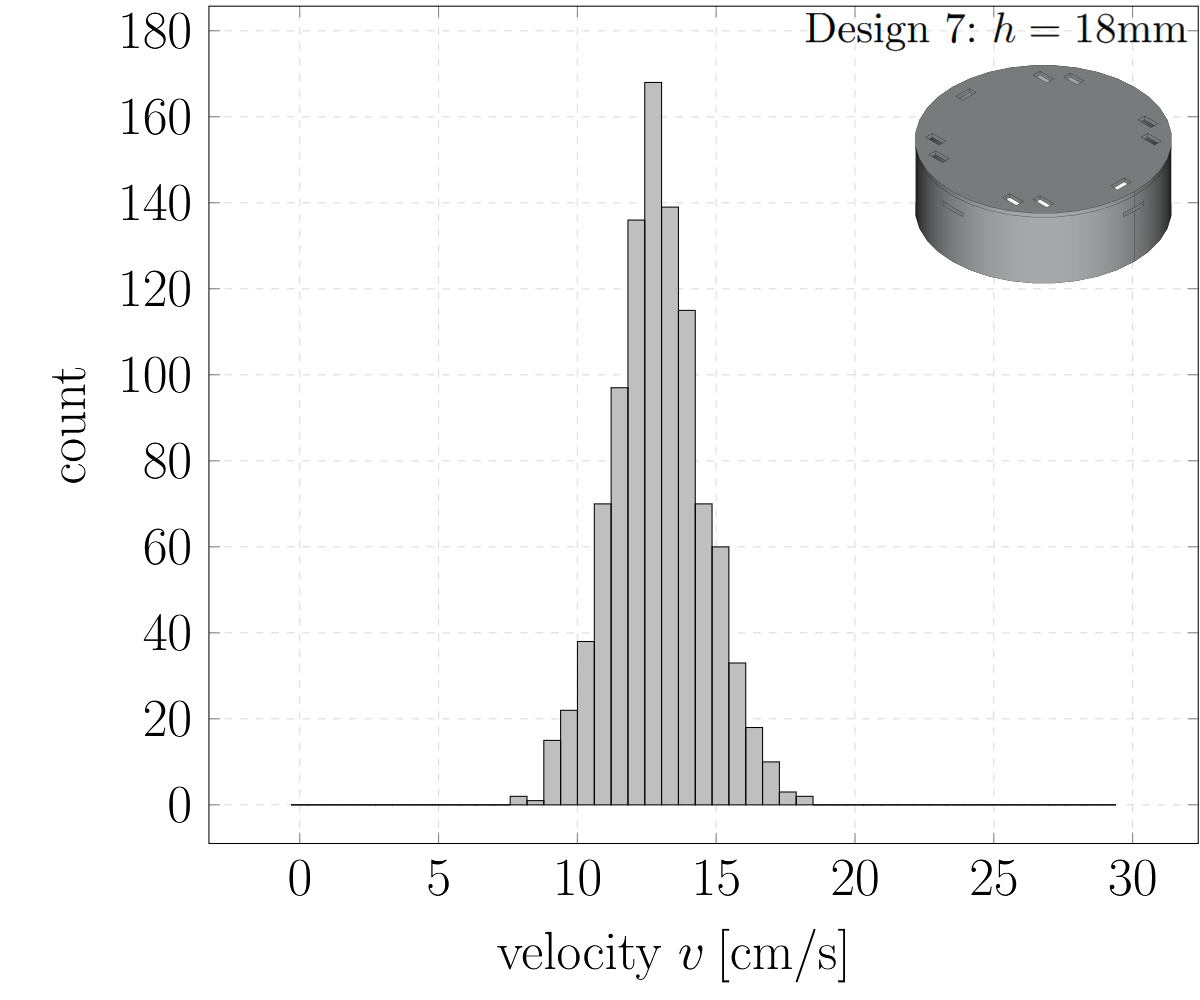}\hfill\includegraphics[width=0.24\linewidth]{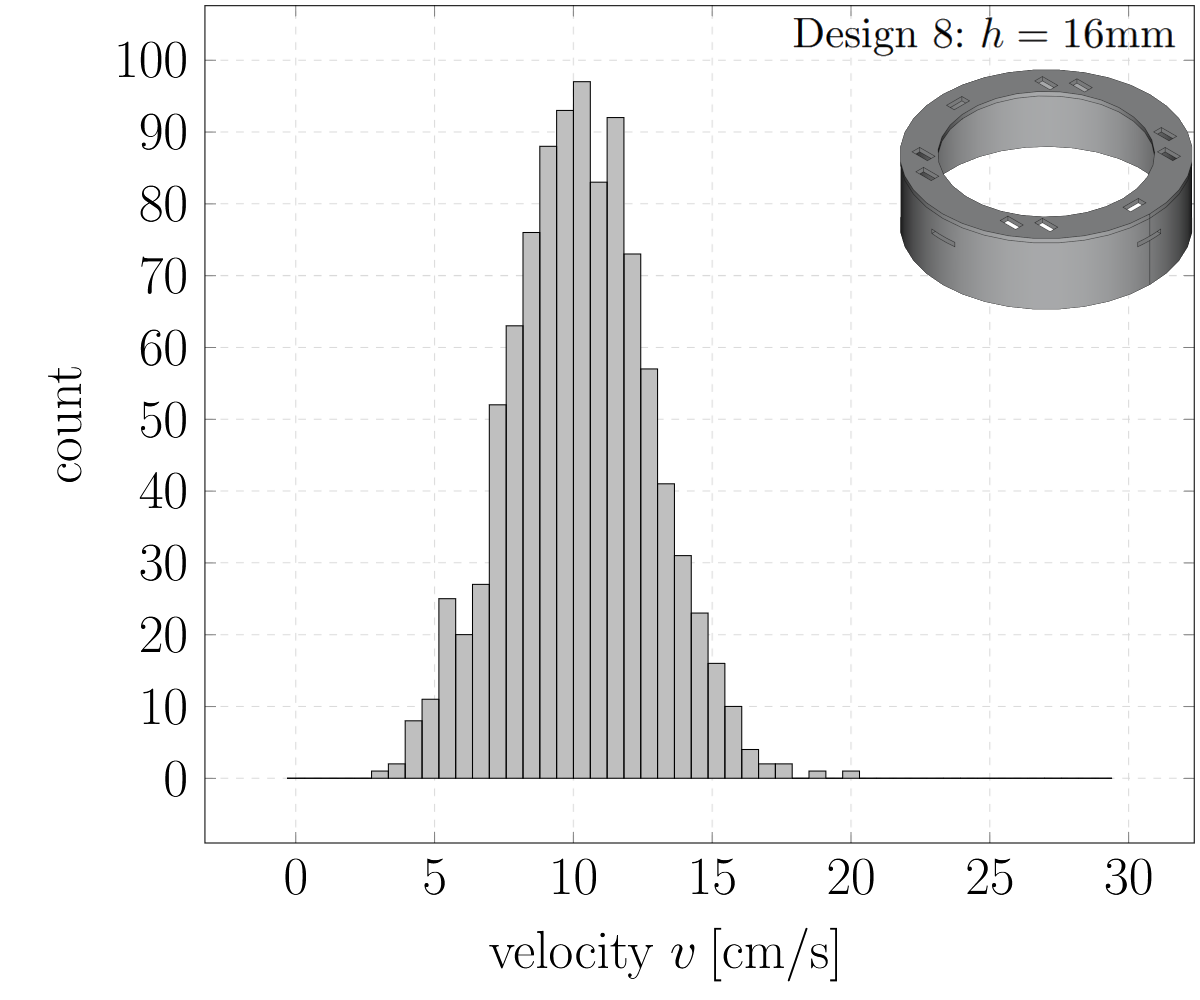}\\
\hfill\includegraphics[width=0.24\linewidth]{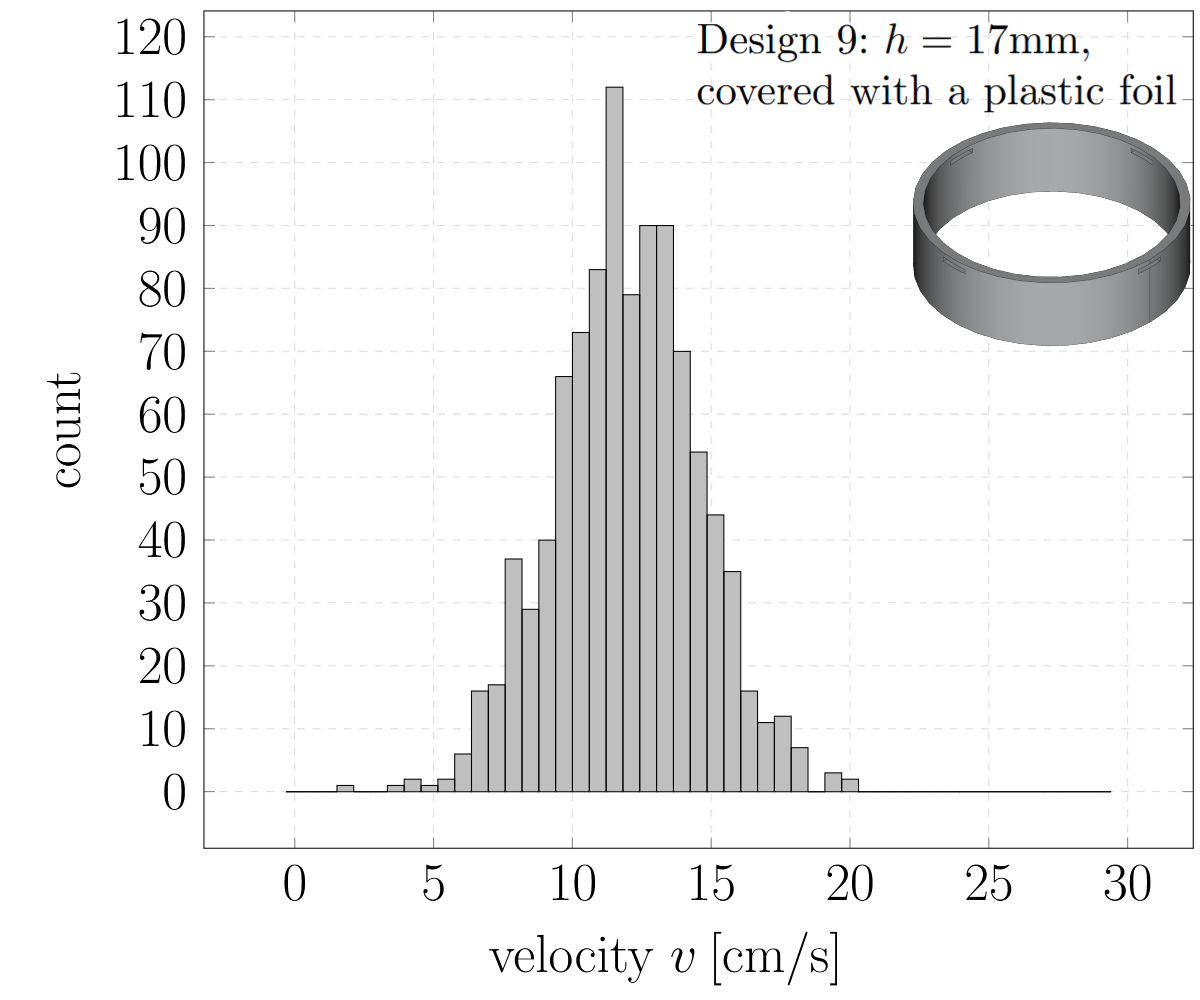}\hfill\includegraphics[width=0.24\linewidth]{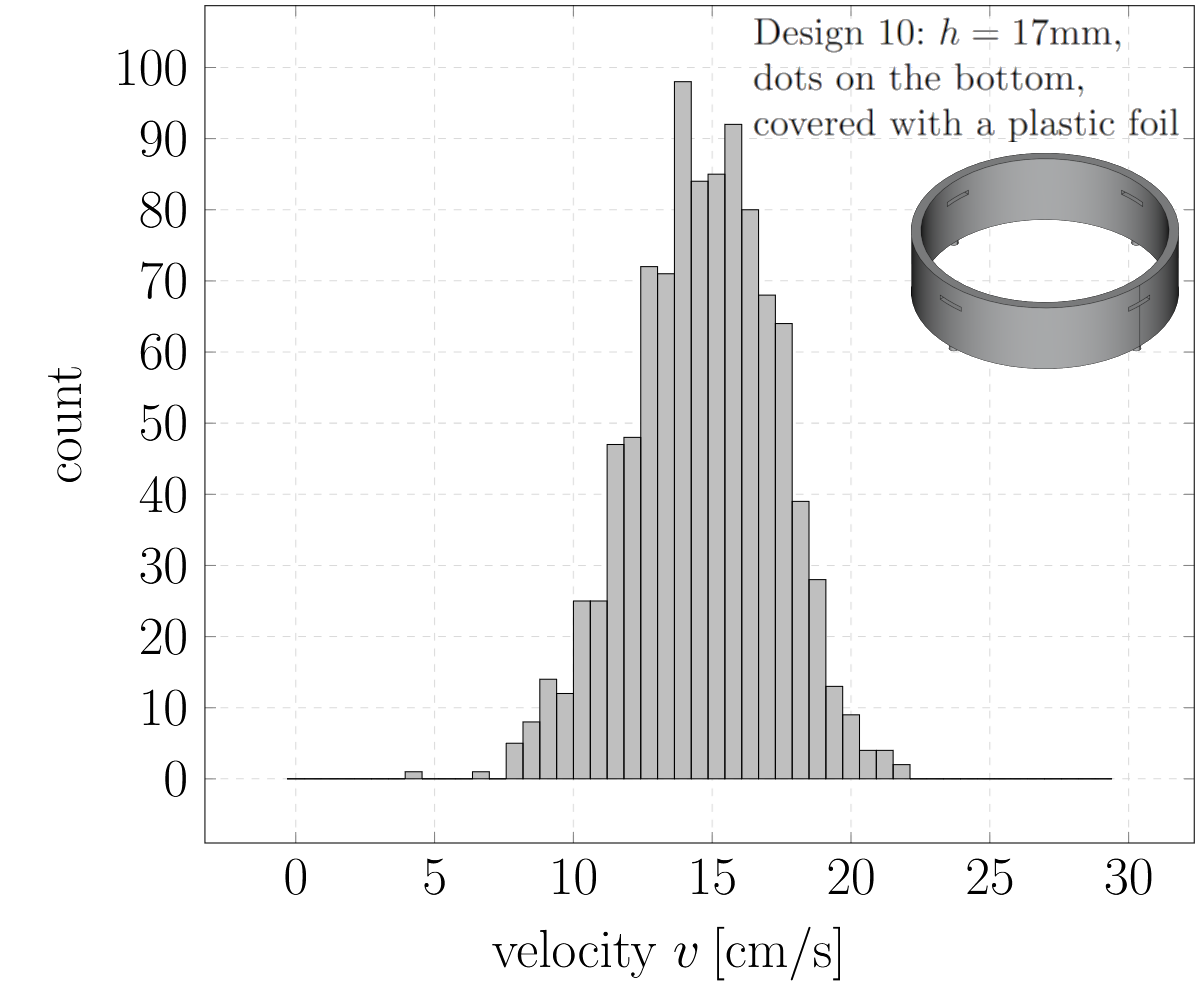}\hfill$~$
\caption{Empirical speed distributions for various versions of the design of the housing. Details of the respective design (as well as a rendering of it) are given as an inset. The last two designs (as well as the final design shown in fig.~\ref{fig:design}) use an elastic cover which is implemented by commercially available foil. Aside from introducing self-alignment to the system, this cover increases the velocity as the static friction~\cite{antonov} is overcome by lifting the housing during the jumps of the nanobots. This is reinforced by tapering the sidewalls (i.e. the shape is not strictly cylindrical, but really a truncated cone).  In designs 2 and 10 the lower rim of the housing was structured by dot-shaped objects to limit the contact area with the ground. }
\label{fig:otherdesigns}\end{sidewaysfigure}
 
\begin{figure}
    \centering
\includegraphics[width=0.5\linewidth]{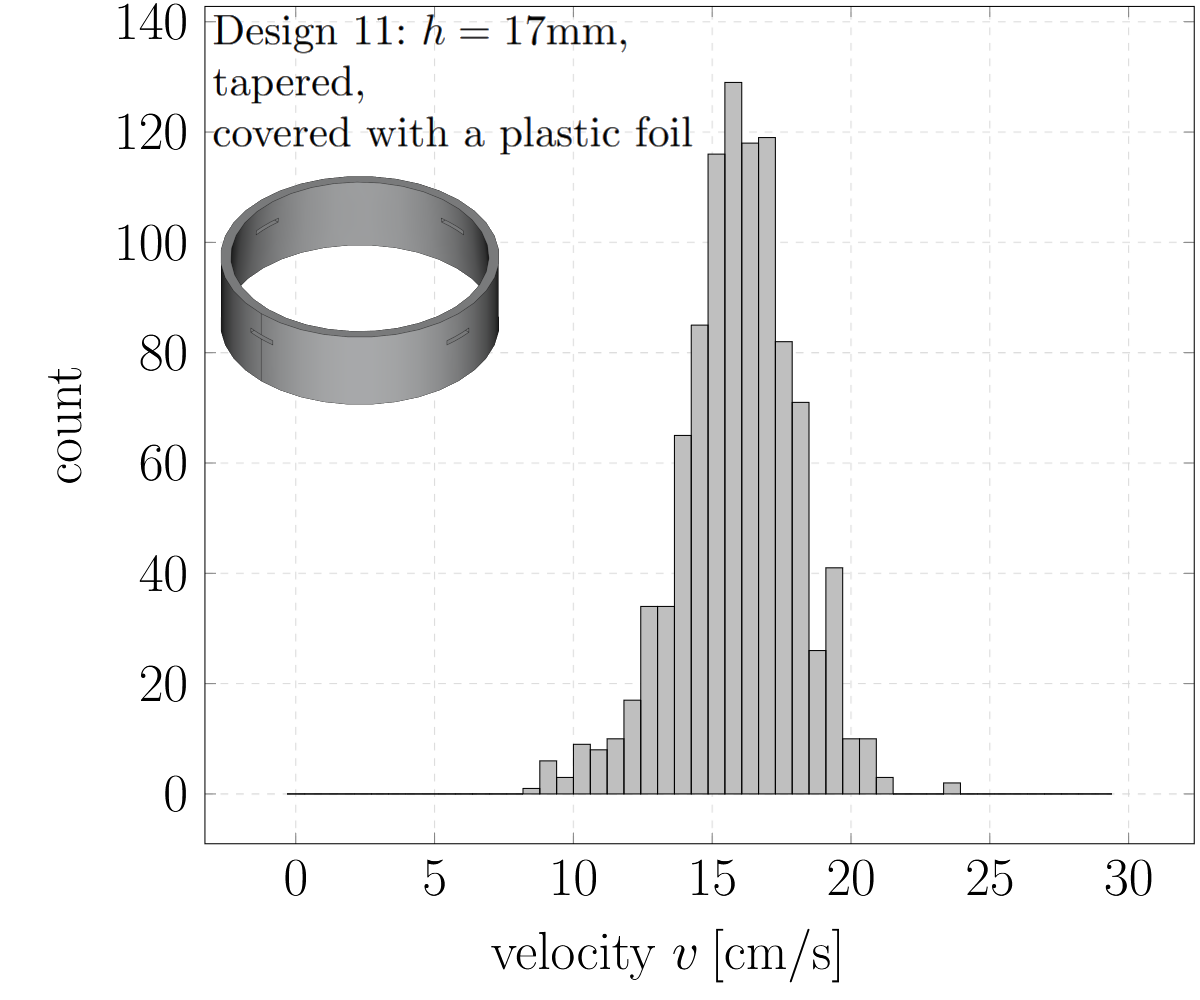}
\caption{Empirical speed distribution for the final design (design 11) that was used in the main text. This design uses the elastic lid, the structured rim and the tapered walls discussed in fig.~\ref{fig:otherdesigns}. We infer $v_0\approx 15.75$cm/s from this distribution. Note that there is a roughly 8-fold increase in the speed from design 1 to this final design, highlighting the sensitivity of this system to simple design choices.}
\label{fig:design}
\end{figure}

We present empirical speed distributions for various housing designs in figs.~\ref{fig:otherdesigns} and fig.~\ref{fig:design}. The design files are available as part of the associated dataset~\cite{zenodo2}. The data clearly demonstrates an enormous increase in the observed speed, greatly increasing the usability in actual experiments not only due to shorter necessary observation times but also because the relative noise strength is decreased and thus there is greater insight into the activity-dominated regime.

This effective reduction of noise has to be balanced against the need for a persistence length that is at most of the order of the arena size and our want of having observable chirality. These aspects are problematic in the version with a fixed lid (which also increases the speed).

\section{ Brownian circle swimmers}
For the sake of self-containedness, we present some analytical results for a simplified version of the discussed dynamics, the Brownian circle swimmer~\cite{vanteeffelen}, that should capture the free dynamics. This is given by
\begin{align*} 
\dot{\vec r} &= v_0 \vec{n}(\phi) \\
\dot{\phi} &= \omega + \sqrt{2D_r}\xi(t)
\end{align*}
with white Gaussian noise $\xi$, $\langle \xi\rangle =0$ and $\langle \xi(t)\xi(t')\rangle = \delta(t-t')$.

\begin{figure}
    \centering
\includegraphics[width=0.4\linewidth]{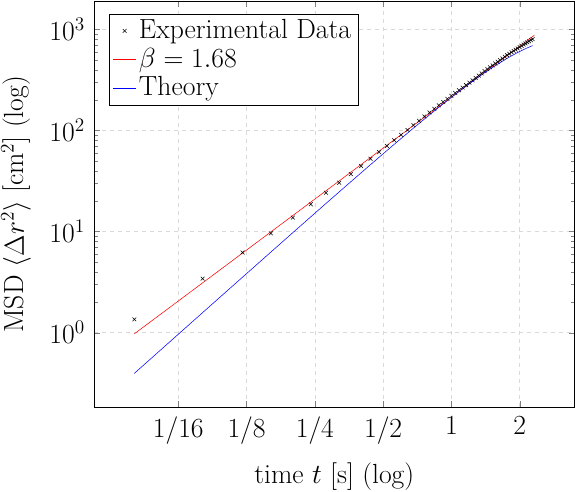}
\caption{Measured mean-squared displacement $\Delta(t)=\langle \Delta \vec r^2\rangle$ shown in a double-logarithmic plot. We find a behavior that is well-described on these time scales by a power-law behavior (red line) $\Delta(t)\sim t^{2a}$ with $2a=\beta\approx 1.68$. However, this is also captured by the Brownian circle swimmer model, see eq.~\eqref{eq:msd} (blue line). There is some deviation due to parameter renormalization from the neglected fluctuations of the speed as well as an additional effective positional noise from particle tracking and the influence of collisions.}
\label{fig:msd}
\end{figure}

We are interested in the velocity correlations $G(t,t')=\langle \dot{\vec r}(t) \dot{\vec r}(t')\rangle$ which by virtue of a Green-Kubo~\cite{gardiner1985} relation also directly give the mean-square displacement (MSD) $\Delta(t)$. Here, we use $\langle \cdot \rangle$ as an average over noise realizations as well as over the initial configurations. Direct insertion gives $G(t',t'') = v_0^2 \langle \cos\phi(t')\cos\phi(t'') + \sin\phi(t')\sin\phi(t'')\rangle$. This is easily computed by expanding the trigonometric functions into complex exponentials and noting that the quantity $q(t)=\phi(t)-\omega t-\phi(0) = \sqrt{2D_r}\int_0^t \xi(t') \mathrm{d}t'$ is a Gaussian stochastic variable with zero mean. From the definition of cumulants and the fact that all higher cumulants of Gaussian variable vanish, we have then $\langle\exp{iq(t)}\rangle= \exp{-D_r t}$. Putting everything together, we find
\begin{align*}
G(t',t'') &= v_0^2 \cos(\omega \lvert t''-t'\rvert) \exp{-D_r \lvert t''-t'\rvert} \text{.}
\end{align*}
Integrating over both times, we get the mean-square displacement
\begin{align}
\Delta(t) &= \frac{2v_0^2}{(\omega^2+D_r^2)^2} \exp{-D_r t} \left[ \exp{D_r t} \{D_r^2(-1+D_r t)+\omega^2(1+D_r t)\} \right. ~+ \left. (D_r^2-\omega^2)\cos{\omega t} -2\omega D_r \sin{\omega t} \right] \label{eq:msd}\\
&\to \begin{cases} v_0^2 t^2 - \frac{D_r v_0^2 t^3}{3} & \text{, for } D_r t \ll 1\\
4 D^* t  & \text{, for } D_r t \gg 1\end{cases} \notag
\end{align}
with a well-known effective diffusion constant $D^* = \frac{v_0^2 D_r}{2(\omega^2+D_r^2)}$. We show that these analytical predictions are indeed consistent with an apparent non-trivial power-law scaling $\Delta(t)\sim t^{2a}$ with $2a\approx 1.68$ that we see on an experimentally accessible scales in fig.~\ref{fig:msd}. Using the analytical predicted correlation functions directly to infer the correlation time-scale (or noise strength $D_r$) is an alternative to the method laid out in the main text. However, we find the latter approach to be better suited to real-life data.

\section{Inter-bot variability}

\begin{figure}
    \centering
\includegraphics[width=0.32\linewidth]{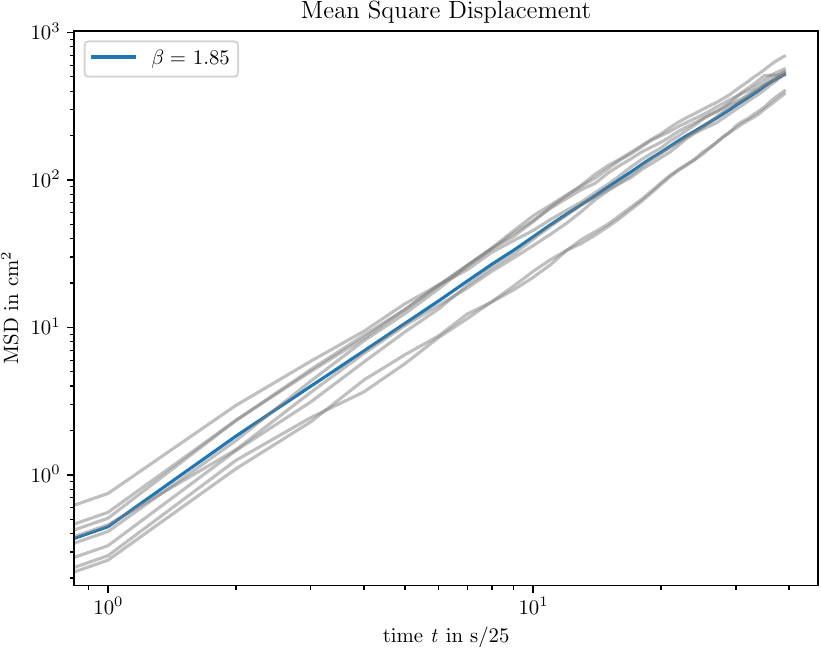}\hfill\includegraphics[width=0.32\linewidth]{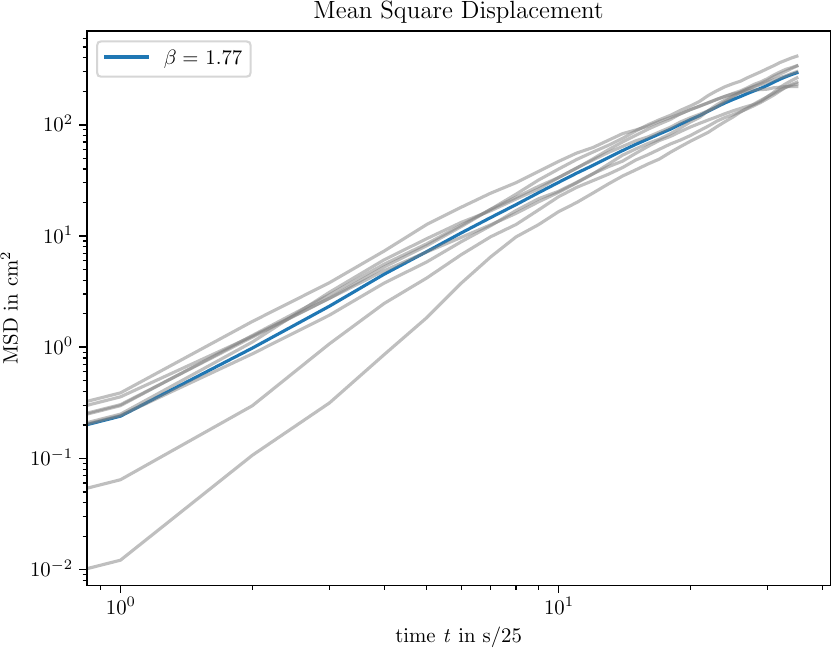}\hfill\includegraphics[width=0.32\linewidth]{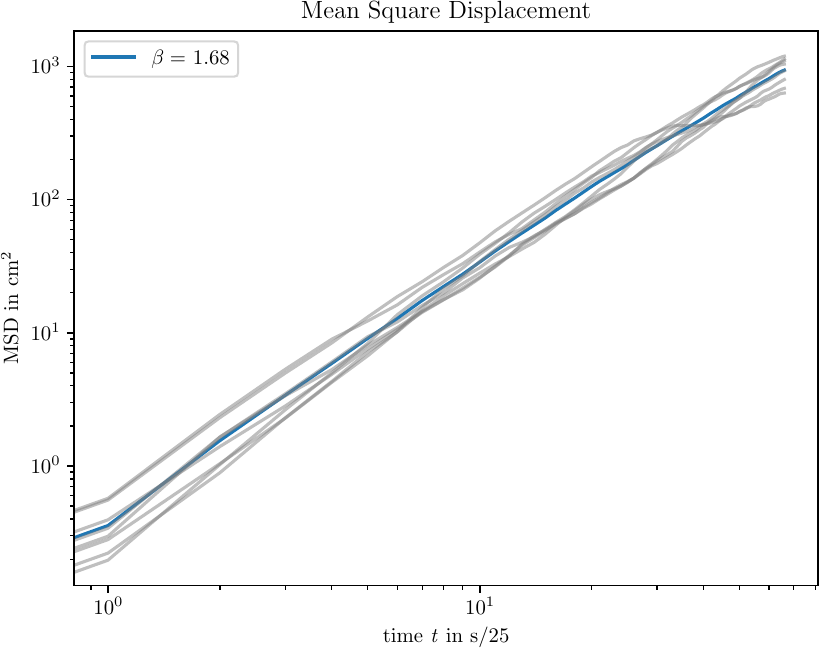}
\caption{Measured mean-squared displacement $\Delta(t)=\langle \Delta \vec r^2\rangle$ shown in a double-logarithmic plot for three exemplary runs. In addition to a power-law fit (blue line) of the entire ensemble as a guide to the eye, we show particle resolved data (gray lines). While we do see some evidence for systematic deviations (center panel, short times), the deviations between experimental runs are more substantial.}
\label{fig:msd2}
\end{figure}

The mean-square displacement data of fig.~\ref{fig:msd}, allows us to analyze the relevance of quenched differences between bots in a bit more detail. As described in the main text, we have chosen 8 bots out of an ensemble of 25 in an attempt to eliminate outliers. We can quantify the effect this has on the ensemble statistics in quick back-of-the-envelope type calculation. Reducing the dynamical behavior to a single one-dimensional feature, one would expect the distribution of that feature to be well described by a Gaussian with some mean $\mu$ and standard deviation $\sigma$. If we indeed are able to identify the 8 particles closest to the mean, we can compare this to truncating the statistics of the distribution to the $8/25=32\%$ closest to the mean. Evaluating the error function shows that this leads to a change in standard deviation $\sigma\to \sigma'\approx 0.24 \sigma$. That is, the elimination of outliers should reduce the variability to roughly a quarter of its original value. While this argument most certainly does not suffice for anything quantitative, it does show the homogenizing effect the outlier rejection has to the studied ensemble. That the inter-bot variability is indeed not a dramatic influence, can be seen by comparing MSD data from various runs of the same experiment, see fig.~\ref{fig:msd2}. We conclude that the inter-bot variability is real and can be relevant in some applications (see main text for some discussion on quasi-1D experiments), we are operating in a setting where other fluctuations (leading to the deviations visible in fig.~\ref{fig:msd}) such as finite experimental duration, battery status, fluctuations in the experimental setup are more important.

\section{Reorientation due to self-alignment}

\begin{figure}
    \resizebox{\linewidth}{!}{
\includegraphics[width=0.52\linewidth]{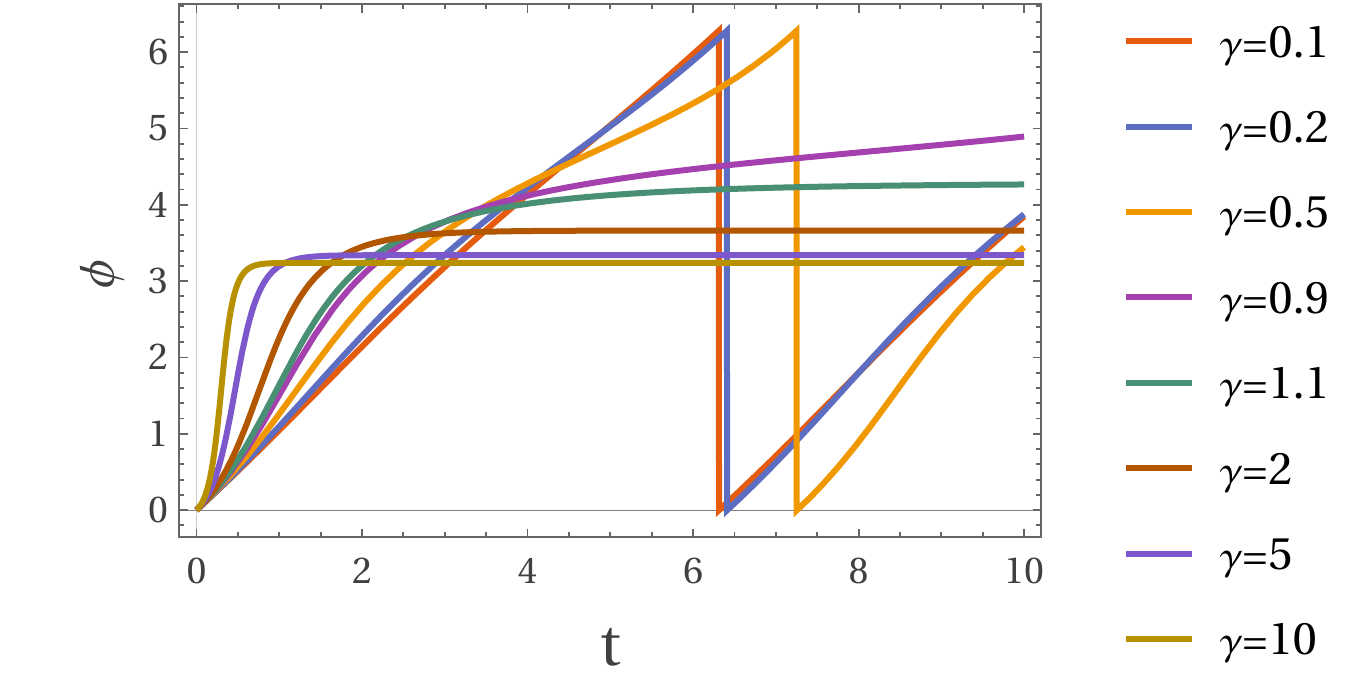}
\includegraphics[width=0.4\linewidth]{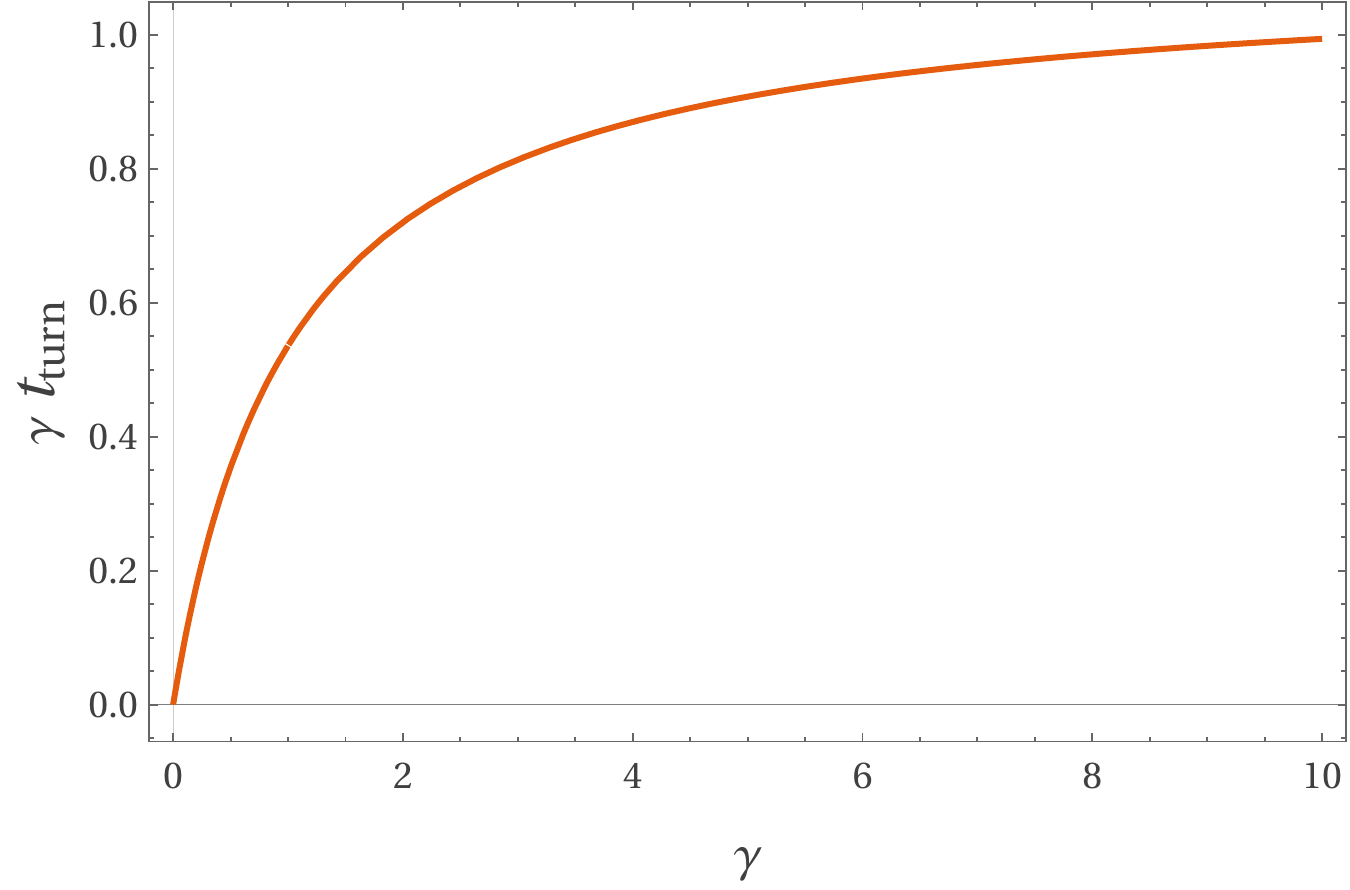}
    }
    \caption{Left: Solutions $\phi(t)$ as given by eq.~\eqref{eq:phioft} for $\psi=0.999\pi$, i.e., the self-alignment under persistent forcing into the direction (effectively) opposite to the initial direction. For $\gamma<1$, the chirality is dominant and the solutions are phase-winding. For $\gamma>1$ as is the case in our experiments where we find $\gamma \approx 4.5$, the solutions show relaxational behavior to a fixed angle (in this case, slightly larger than the externally set angle, $\phi=\psi+2\operatorname{atan}(\gamma-\sqrt{\gamma^2-1})\sim \psi+1/\gamma$, as we are turning with the chirality). Right: We compute the time $t_{\text{turn}}$ it takes to deterministically reorient from $\pi/3$ to $2\pi/3$. The idea is to eliminate the parts of the curve with rather small slope $\partial_t \phi$ as the noise would be the dominant driver there. The specific criterion is somewhat arbitrary, but does not change the result qualitatively. Dimensionally, one expects $t_{\text{turn}} \sim 1/\gamma$. We show that this is essentially correct, but there are some subleading corrections as the product $\gamma t_{\text{turn}}$ is only asymptotically constant. Thus, accurate inference of the self-alignment $\gamma$ is non-trivial, which is enhanced by the noise-driven fluctuations that are not considered here. However, the relevant quantity for the physics is that $\gamma>1$, which is easily confirmed.}
    \label{fig:angle}
\end{figure}

In the case of a persistent driving of the augmented particle's velocity as is for example the case in the experiment used to infer the self-alignment strength (cf. the file \textsc{orientation.mp4} as well as ref.~\citenum{baconnier2022}), the relevant deterministic angle dynamics are in the same non-dimensional units the main manuscript uses given by the Adler equation~\cite{adler}
\begin{align*}
\partial_t \phi &= \gamma \sin(\psi -\phi) + 1 \text{.}
\end{align*}
Herein, $\psi$ is a fixed parameter. Without loss of generality, we can set that $\phi(0)=0$ and then find the solution via Weierstraß substitution to be
\begin{subequations}\label{eq:phioft}
\begin{align}
\phi(t) &= \psi -2 \operatorname{atan}\left[ -\gamma + \sqrt{1-\gamma^2} \tan{\frac{-\sqrt{1-\gamma^2}\,t-2 q_1}{2}} \right] \\
&= \psi - 2 \operatorname{atan}\left[ -\gamma + \sqrt{\gamma^2-1} \tanh{\frac{\sqrt{\gamma^2-1}\,t+2q_2}{2}} \right]   
\end{align}
\end{subequations}
with
\begin{align*} 
     q_1 &= \operatorname{atan} \frac{\gamma \sqrt {1-\gamma^2} + \sqrt{1-\gamma^2}\tan{\frac{\psi}{2}}}{1-\gamma^2} \qquad \text{ and} \\
     q_2 &= \operatorname{artanh}\frac{\gamma  + \tan{\frac{\psi}{2}}}{\sqrt{\gamma^2-1}},
\end{align*} respectively. The two presentations of this result are equivalent, but highlight that the nature of these solutions changes drastically between $\gamma>1$ and $\gamma<1$ as it changes which of the occurring roots are real. We show this graphically for $\psi=0.999\pi$ (effectively corresponding to a full turn avoiding the unstable equilibrium for $\phi(0)=0$, $\psi=\pi$ for which fluctuations would always be relevant) in fig.~\ref{fig:angle}.
 
In the case of very large self-alignment couplings $\zeta\gg\omega$, it is no longer useful to apply the rescaling of time in terms of $\omega^{-1}$ instead one should consider the effective dynamics
\begin{align*}
\partial_t \phi &= \sin(\psi -\phi)  
\end{align*}
directly. This is readily solved to give $\phi_{\zeta\gg \omega,\text{det}}=\psi + \operatorname{acos}\tanh\left[t + \operatorname{artanh}\cos{(\phi(0) - \psi)}\right]$. In fact, we can also solve the full stochastic problem
\begin{align}
\partial_t \phi &= \zeta \sin(\psi -\phi) + \omega + \sqrt{2 D_r} \, \xi \label{eq:eom}
\end{align}
in the sense of determining the stationary distribution~\cite{stratonovich,risken}. While this is a well-discussed problem in the literature, we give some details here for self-containedness. We note that the deterministic part is of the form of a gradient flow, i.e., $\partial_t \phi = - V'(\phi) + \omega \sqrt{2 D_r} \, \xi$ with the ``washboard'' potential $V(\phi) = - \omega \phi - \zeta \cos(\psi-\phi)$ that is a staple of nonequilibrium statistical physics and has seen successful application in the DC Josephson effect. The Fokker-Planck equation for the probability distribution $P(\phi,t)$ of $\phi$ is then
\begin{align*}
\partial_t P &= \partial_\phi \cdot (V' P + D_r \partial_\phi P) = -\partial_\phi J 
\end{align*}
with a probability flux $J$. As the probability is overall a conserved quantity, the Fokker-Planck equation has to take the form of a continuity equation. We are interested in the stationary case, which corresponds to $\partial_t P \equiv 0$. In equilibrium, the flux vanishes, $J\equiv 0$, which is reflective of detailed balance, and the solution is the Boltzmann weight $P_{eq}\propto \exp{-V/D_r}$. Here, we are out of equilibrium due to the drift $\omega$ and $J\equiv J_0$ takes a constant non-zero value. However, we can directly see that the Boltzmann weight acts as an integrating factor, i.e. we write $P=P_{\text{eq}}\, Q$, and see
\begin{align*}
P' +  \frac{V'}{D_r} P &= - \frac{J_0}{D_r} \\
Q' &= - \frac{J_0}{D_r} P_{\text{eq}}^{-1}
\end{align*}
and, thus, we find (using periodic boundary conditions) the classical solution
\begin{align*}
P(\phi) & \propto  \exp{-V(\phi)/D_r} \int_\phi^{\phi+2\pi} \exp{V(\theta)/D_r} \,\mathrm{d}\theta \text{.}
\end{align*}
From this, we find $J_0 = D_r (1-\exp{-2\pi\omega/D_r})$ and  can, for example, compute the mean drift in the nonequilibrium stationary state~\cite{Ambegaokar} by averaging eq.~\eqref{eq:eom} with the stationary distribution
\begin{align*}
\langle \dot \phi\rangle&= \frac{2\pi D_r (1-\exp{-2\pi\omega/D_r})}{\int_0^{2\pi} \mathrm{d}\phi\, \int_{\phi}^{\phi+2\pi}\mathrm{d}\theta\, \exp{\frac{V(\theta)-V(\phi)}{D_r}}} \text{.}
\end{align*}
Finding a dynamical solution~\cite{cheng2018} for the probability distribution is, however, a more challenging task that requires more advanced (numerical) techniques, such as an expansion into the eigenfunctions of the Fokker-Planck operator which often involves mapping to a Schrödinger equation~\cite{steinhoefel}. In many applications, the short-time behavior is of interest as the forcing direction $\psi$ is subject to stochastic dynamics on its own. We can directly find this as a drifting Gaussian from the Fokker-Planck equation
\begin{align*}
P(\phi,t \vert \phi_0,0) &\approx \frac{1}{\sqrt{4\pi D_r t}} \operatorname{exp}{\left(- \frac{\left[\phi -\phi_0 -(\omega +\zeta \sin(\psi-\phi_0)\,t)\right]^2}{4 D_r t}\right)}    \text{.}
\end{align*}

\section{Wall residence times}

As noted in the main text, the bots form rather stable currents at the edge of the circular arena. To substantiate the statistical analysis of the radial distribution function, we also consider the wall residence times as a dynamical quantifier of the wall interaction. We define these as the time for which particles are within a fixed distance $\delta$ of the edge of the circular arena. As noted before, there is an important distinction to be made with respect to the handedness of the motion at the wall which manifests in the effective interaction with the wall and therefore the distribution of residence times near it.

\begin{figure}
    \includegraphics[width=.32\linewidth]{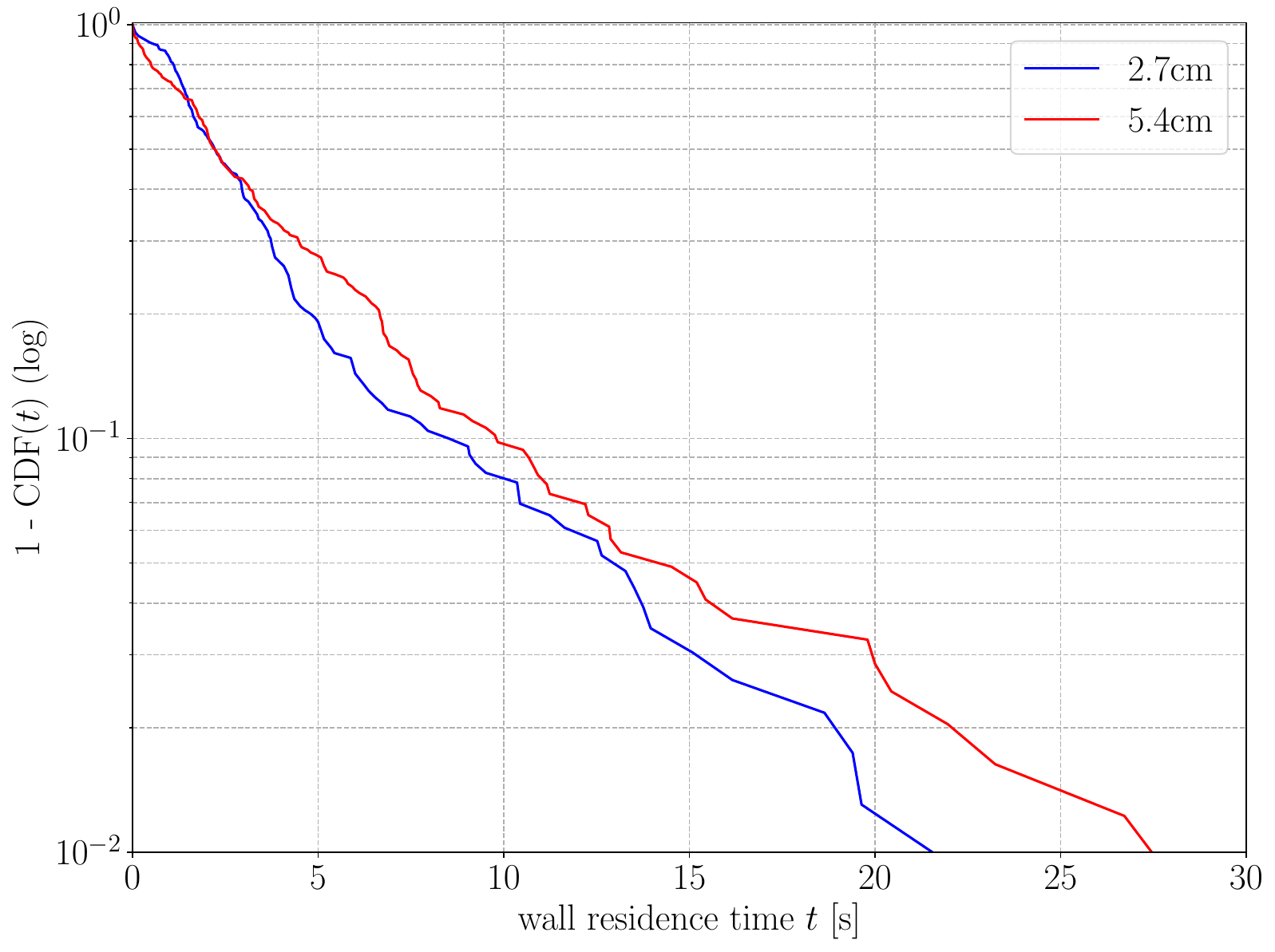}\hfill
        \includegraphics[width=.32\linewidth]{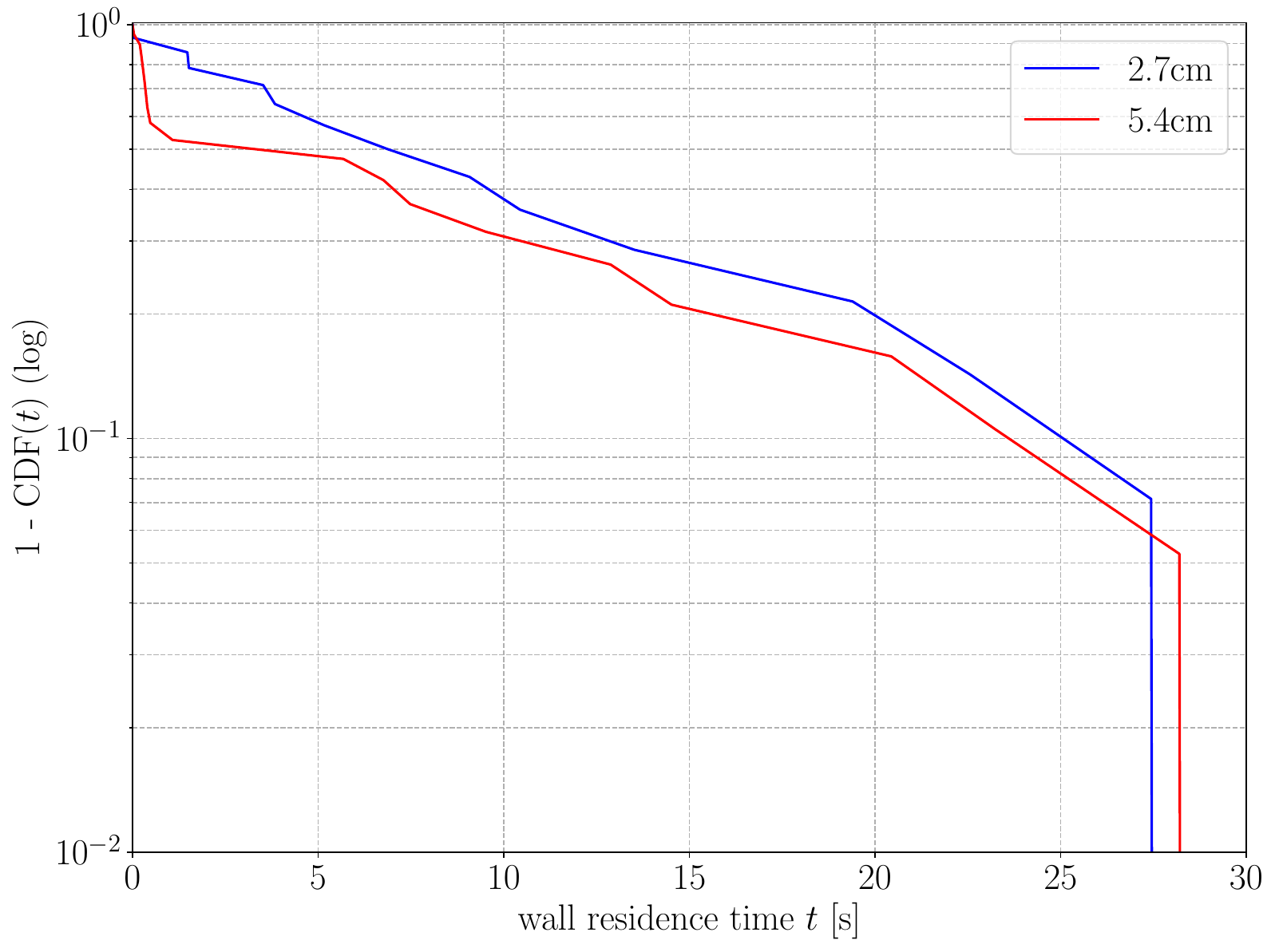}\hfill
            \includegraphics[width=.32\linewidth]{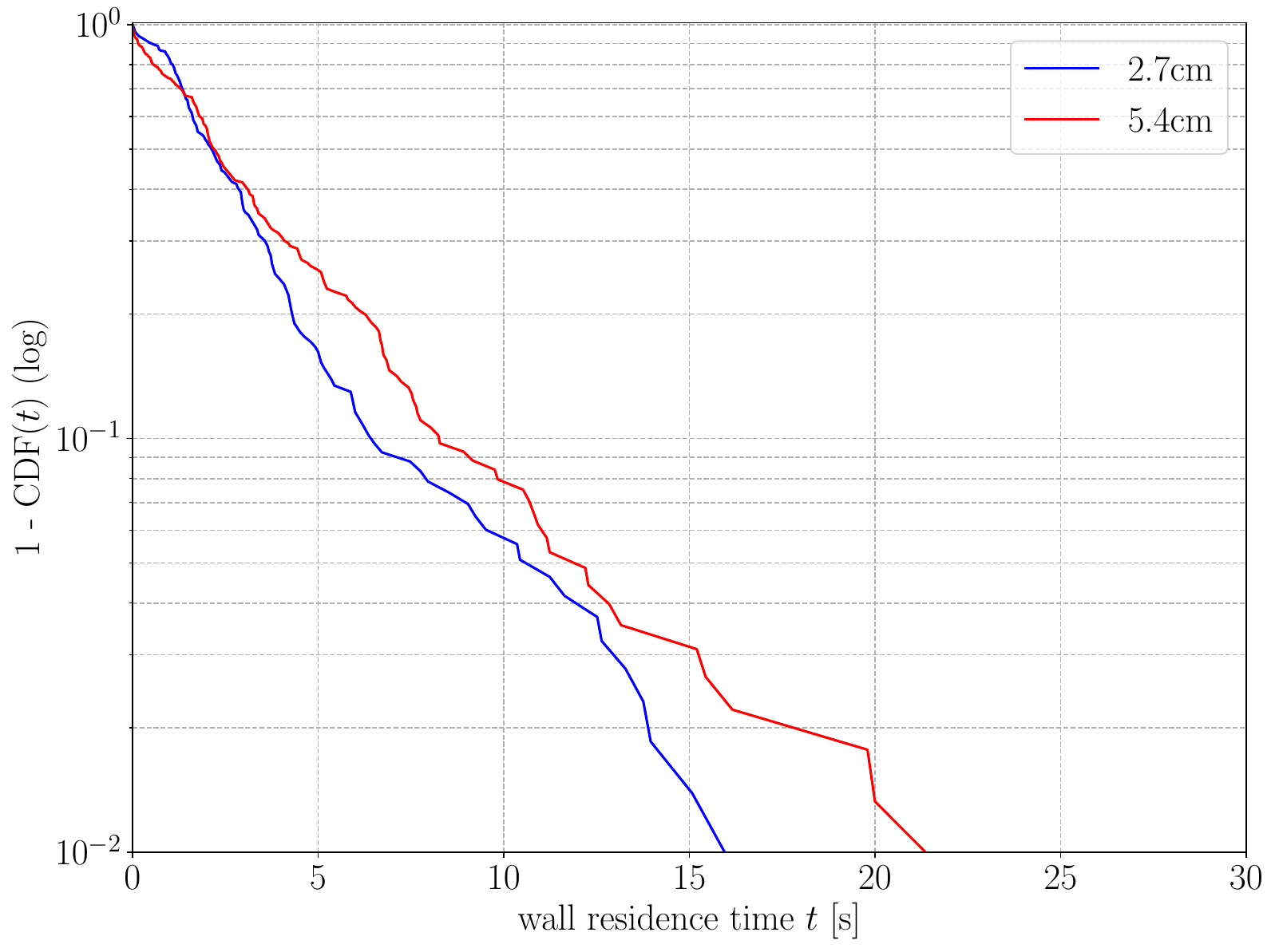}
    \caption{Distribution of wall residence times for two different distances $\delta$. We consider $\delta=\ell_h=2.7$cm and $\delta=2\ell_h=5.4$cm and see only a weak dependence on this parameter. Shown are the survival functions $1-\mathrm{CDF}(t)$ wherein $\mathrm{CDF}(t)$ is the empirical cumulative distribution function for the residence time. This analysis circumvents the need for binning. As we expect the times to be governed by an Arrhenius-type law with some timescale $\tau_{W}$, i.e. $1-\mathrm{CDF}(t)\approx \exp{-t/\tau_{W}}$, we show the data using semilogarithmic axes. Left: All trajectories. Center: Only trajectories, where the motion close to the wall is net counter-clockwise. Right: Only trajectories that are net clockwise.}
    \label{fig:residencetime}
\end{figure}

In Fig.~\ref{fig:residencetime}, we present the data for two values of $\delta$ corresponding to one housing radius and one housing diameter, respectively. The latter corresponds to the first next ``shell'' visible in the radial distribution function, which serves as a natural cut-off condition to the amount of wobble the motion near the wall can sustain. Taking all trajectories into account (left panel), we see two time-scales: the survival function decays rather fast (timescale $\tau_{W,\text{fast}}\approx 3$s) on short times, but at larger times a slower decay is observed (timescale $\tau_{W,\text{slow, app}}\approx 6-7$s). Remarkably, this is rather independent of the choice of $\delta$. Splitting the trajectory segments close to the wall by whether the motion from the start- to end-point corresponds to a clockwise or counterclockwise spatial motion, we can mostly resolve the origin of this. The residence time is indeed found to be heavily dependent on the handedness of the motion, with counter-clockwise motion leading to very long (center panel) wall times with a typical timescale $\tau_{W,\text{slow,ccw}}\approx 11$s and clockwise motion to shorter times (right panel). Additionally, the counter-clockwise motion does display a sensitivity in $\delta$. The data for the wider gap shows a drastic decay at short times, these correspond to turning induced by the deterministic chirality $\omega$.

\section{Positional Statistics in Geometric Rectification}

\begin{figure}
    \includegraphics[width=\linewidth]{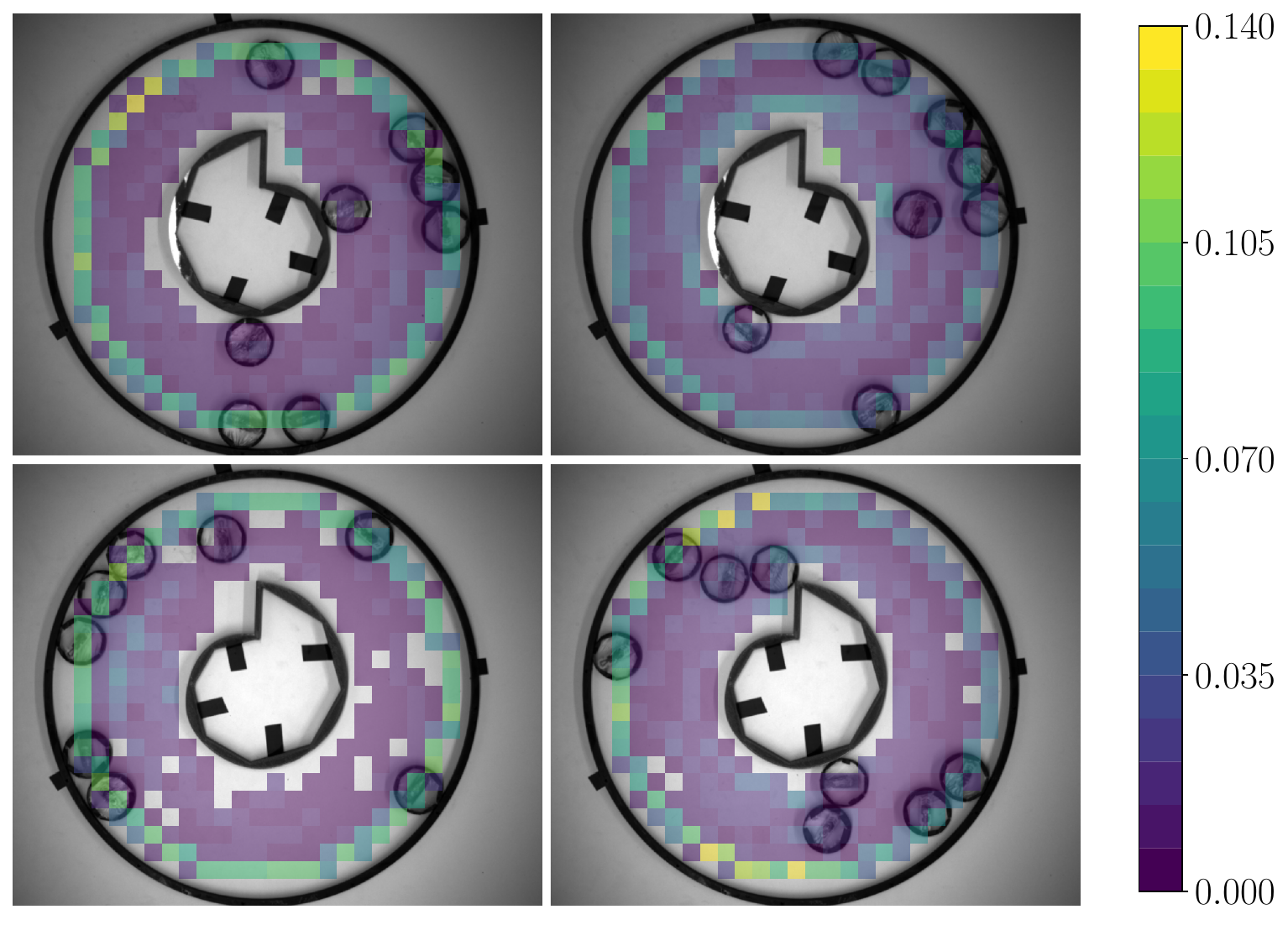}
    \caption{Larger visualization of the spatial statistics of the bots in the chiral environment under the four handedness conditions discussed in the main text. Throughout all panels it is apparent that there is an excess aggregation at the outer border, which is consistent with the observations without the central obstacle. Unfavorable orientations (most clearly lower right) lead to a relative overpopulation of the corner of the nautilus shape, which as the only locally convex section of the geometry is the predominant spot for aggregation. Particles that move into this spot, will stay there for a while.  In the least-impeding case (lower left), the edge current is very stable and any regions that would be unfavorable to transport appear severely depleted.}
    \label{fig:heatmap}
\end{figure}

In Fig.~\ref{fig:heatmap}, we present the heatmaps that are presented as an inset in the main manuscript. To quantify, where the geometrically induced slowing-down observed in the velocity histograms is taking place, we compute empirical coarse-grained spatial densities. These are then visualized using a heatmap that is consistent throughout the panels.

Here, the depletion near the ``corner'' of the nautilus is the most relevant quantity. In the case of most favorable transport (lower left), particles being there is so unlikely that we cannot show data for it, whereas there is an accumulation in the least favorable configuration (lower right).  Additionally, one observes the expected formation of a second (partial) edge current in the configuration in the upper-right.

\section{Motility induced phase separation}

\begin{figure}
\includegraphics[width=\linewidth]{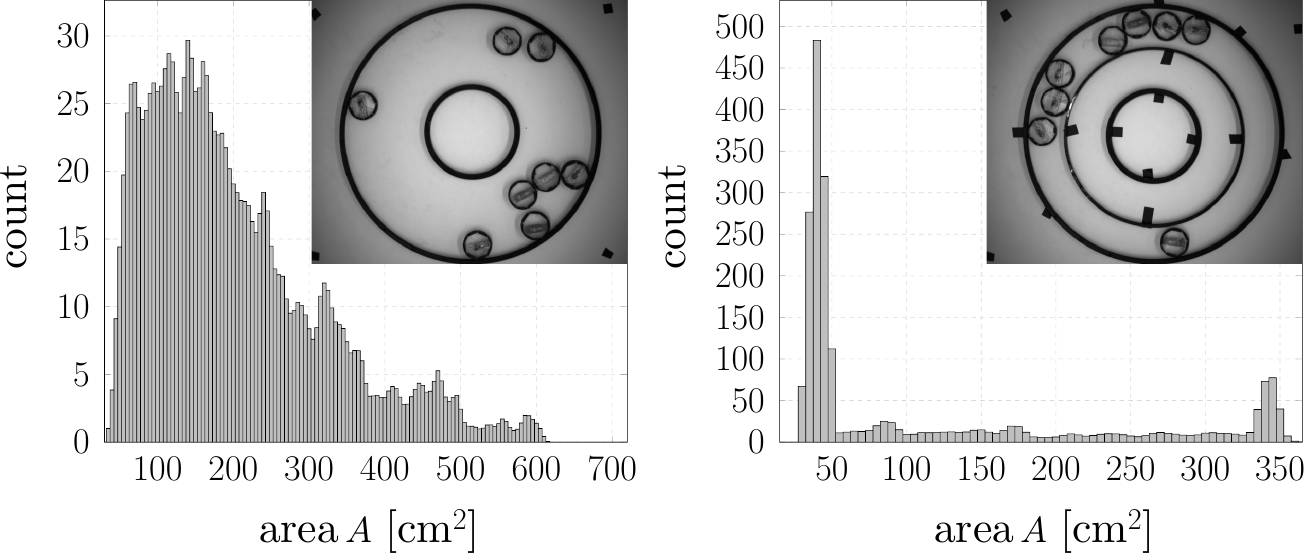}
\caption{Empirical distribution of areas per particle by means of Voronoi tesselation~\cite{knippenberg} for two scenarios in which overtaking is difficult due to hard constraints in contrast to the softer constraint~\cite{knippenberg} in fig.~\ref{fig:6}. Left: Roughly two-lane traffic with some indication for excess statistical weight at large areas (low densities).  Right: One lane traffic with a rather clear separation of the system. However since overtaking is classically not possible in this scenario, the collective transport is limited by the slowest particle and, therefore, this is not a collective effect. In transients, more genuine traffic jam phenomenology can be observed wherein the separation is induced by direction/speed fluctuations, but the separation is nucleated consistently at the slowest particle in longer runs.}
\label{fig:mips1}
\end{figure}

\begin{figure}
\includegraphics[width=\linewidth]{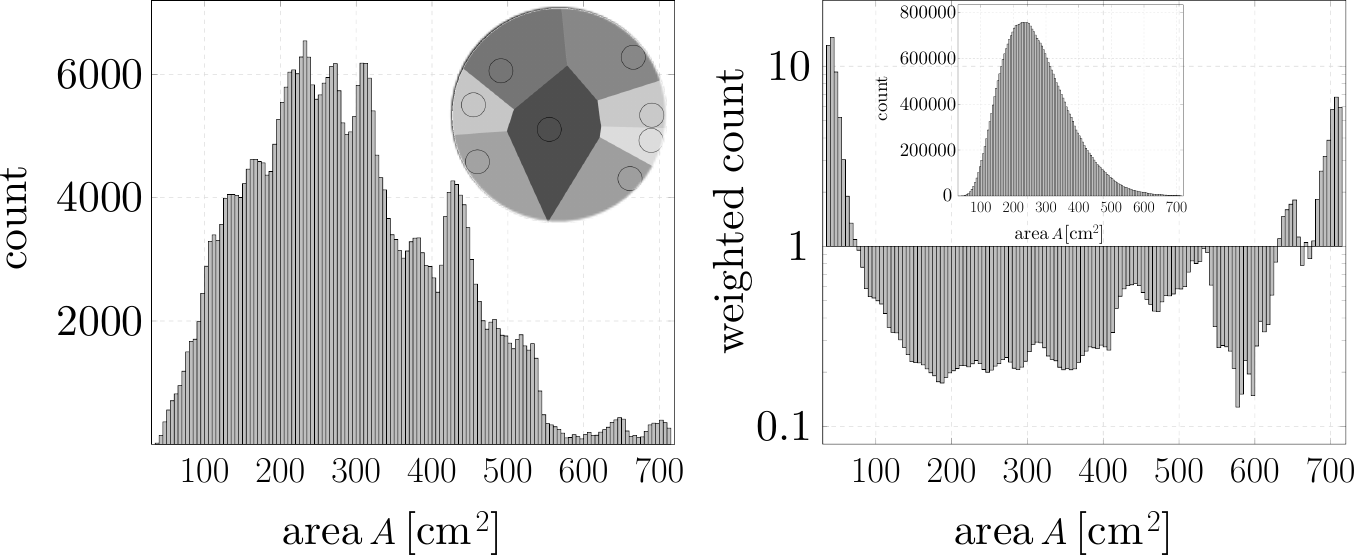}
\caption{Left: Empirical distribution of areas per particle by means of Voronoi tesselation (example shown as inset, shading represents area). As the dynamics favors particles to be close to the wall, this can be seen as a soft constraint to the interior direction, while the exterior direction is a hard wall~\cite{knippenberg}.  Right: Relative count of areas compared to synthetic data (randomly sampling non-overlapping circles, see inset) highlighting the separation into high density (low area) and low density (high area) statistics.}
\label{fig:6}
\end{figure}

We observe a wall-nucleated version of motility induced phase separation (MIPS) in our system. The two phases are: a free chiral gas in the interior at rather low densities and a chiral flock close to the wall at high densities.  As our systems are not large enough to meaningfully define mesoscalic densities, we take an approach~\cite{knippenberg}
 based upon Voronoi tessellations to assign a local area (and hence a density as its inverse) to every particle. When we constrain the motion of the particle such that overtaking is difficult or impossible, we find a clearly observable separation phenomenology, see fig.~\ref{fig:mips1}. In our case, this has to be attributed to quenched differences in particle speed and, thus, is not a collective effect. These quenched differences are not as relevant in the case of an open circular arena. For this case, we show our methodology in Fig.~\ref{fig:6}. By comparing the area distributions from Voronoi tessellations (constrained by the circular boundary) of the experimental trajectories (left) to synthetic data (inset right, generated by placing random points with the correct non-overlapping constraint), we find a clear excess of very small and very large areas which corresponds to an overabundance of large and small densities. 

 \section{In-silico representation}

\begin{figure}
\includegraphics[width=\linewidth]{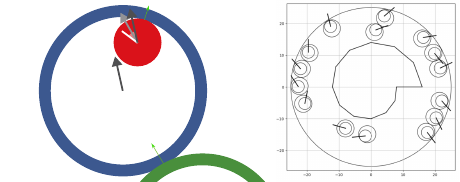}
\caption{Left:  Sketch of the numerical model. An active particle (red) is placed within an external ring (blue). It has an internal direction (white bar) along which it is self-propelled. There are contact forces (green dashed arrows) acting on the ring (corresponding counter-forces not shown) from the internal active particle as well as external objects (other rings, walls), giving it a velocity (dark gray). The steric forces on the active particle lead to its velocity (light gray) differing from the direction of self-propulsion. Additionally, there is a torque on this direction aligning it with the motion of the external ring (indicated by a shaded area between the propulsion direction a copy of the ring velocity). For visual clarity, the individual parts are not to scale.  Right: Exemplary snapshot reproducing the experiment with a nautilus-shaped chiral object in the center of the arena. Each augmented hexbug is represented by an internal circle (the bristlebot) and an external ring (the housing) as well as straight line indicating the direction of the self-propulsion force. We simplify the evaluation of wall forces by reducing the geometry to circles and straight lines.}
\label{fig:simulation}
\end{figure}

 The simple structure of our experiments makes it possible to formulate a numerical representation of it that adequately captures the main aspects. Especially, we can include the internal flexibility (position of the driving bristlebot within the housing) that is harder to treat analytically. Our approach is to reduce the system to simple particles that interact by way of Hertzian contact mechanics. The bristlebot is represented as a circular particle placed within a ring-shaped particle (representing the housing). The relevant details of the motion: self alignment and chirality are directly implemented into the motion of this particle, it couples to the direction of motion of its respective housing and has a drift. This outer particle interacts with other housings and the walls implementing the steric repulsion.  The force constants of the contact interaction (the elastic moduli) are in principle accessible but not really relevant as the simulation only requires them to be sufficiently large that any overlap is only present for very short times. All other parameters are either directly accessible or can be gauged by means of the same quantities as the experiment leading to a numerical model without any meaningful free parameters. Every augmented-bristlebot particle is thus described by the positions of the center-of-mass of the ring $\vec r_i^r$ and bristlebot $\vec r_i^b$ as well as an internal angle $\phi_i$ that encodes the direction of the bristlebot. We can write their overdamped dynamics as
 \begin{align}
 \dot{\vec r_i^r} &= \mu \vec F^{rb}_i + \sum_{j\neq i}\vec F^{rr}_{ij} + \vec F^{rw}_i \propto (\cos \psi_i, \sin \psi_i)^{\mathrm{T}}\\
 \dot{\vec r_i^b} &= -\vec F^{rb}_i + v_0 \vec{n}(\phi_i) \\
 \dot{\phi_i} &= \zeta \sin(\psi_i-\phi_i) + \omega+\sqrt{2 D_r}\xi_i
 \end{align}
 wherein $\vec F^{ab}$ are contact forces due to interactions of objects of types $a$ and $b$ (with $r$ for ring, $b$ for bristlebot and $w$ for external walls) and the self-alignment mechanics is similarly realized as in the main manuscript keeping in mind that the elastic coupling (exerting a torque on the bot) depends on the ring velocity. In the case that the ring is not moving, we do not evaluate the self-alignment. The parameter $\mu$ reflects that we consider overdamped dynamics and the mobilities of bot and housing are likely not identical, thus Newton third's law cannot be presumed to hold. In the other steric forces, this distinction is not material as the force strengths for ring-ring and ring-wall interactions only affect the ring's motion. This is a simplification as the ring is in reality moved due to being lifted from jumps of the bristlebots, but the relevant elements are captured. We show some proof-of-concept realization in fig.~\ref{fig:simulation} and will explore this further in future work.

\begin{figure}
\includegraphics[width=.49\linewidth]{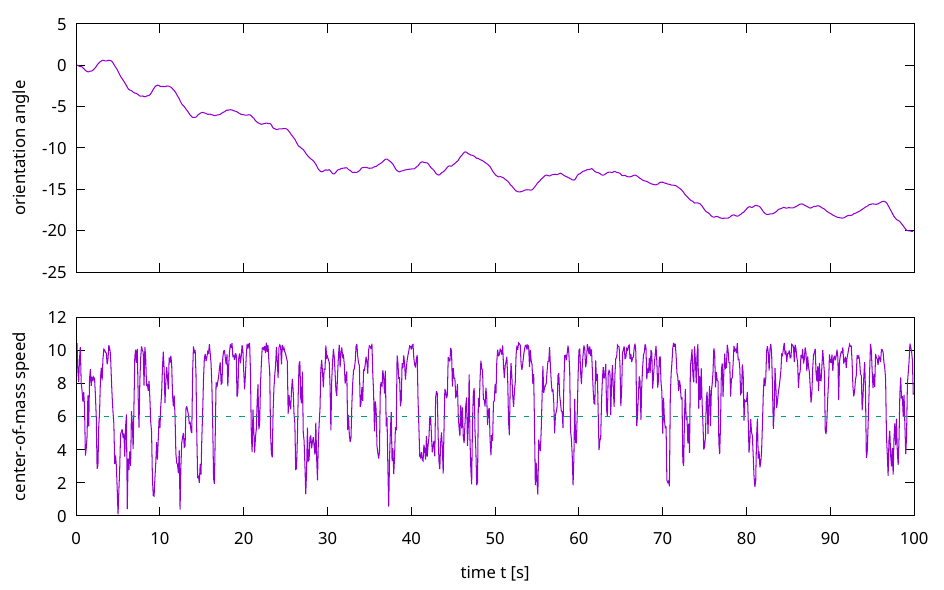}\hfill\includegraphics[width=.249\linewidth]{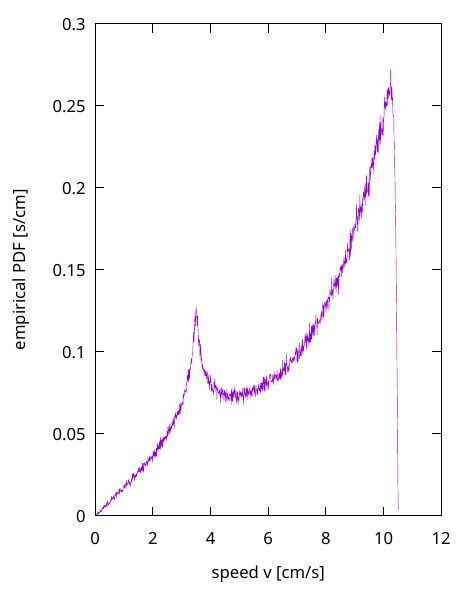}\hfill\includegraphics[width=.249\linewidth]{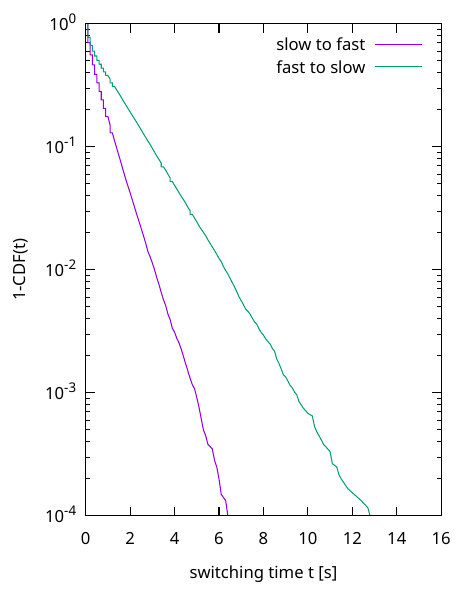}
\caption{Results from numerical simulation of the triangle-constrained assemblage of self-aligning chiral bots. Left: Time trajectories of a rather short (yet substantially longer than any experiment) run depicting the orientation angle (top) and center-of-mass speed (bottom) of the triangle.  The horizontal line in the lower panel denotes an ad-hoc threshold in the speeds that was used to identify the two collective states. This is based on the empirical probability density function (center panel).  Having introduced a definition of states, we can measure the switching times, we show empirical cumulative density functions (right panel) showing a very clear Poissonian behavior. Furthermore, we did not find any evidence for correlations between subsequent times.}
\label{fig:simulation2}
\end{figure}

As an exemplary application of the numerical model, we consider the case of three bots with the rigid constraints holding the housings in an equilateral triangle, as considered in the main text. Tuning the simulation parameters to values that reproduce the gauge measurements (chirality, self-aligning, overall speed), we find little to no evidence of switching. However, the mass of the housings is effectively larger with the constraints so that $\mu$ is unclear. Additionally, there can be an effectively larger noise and reduced (friction-induced) self-aligning due to a synchronization of the jumps. Therefore, a larger region of the parameter-space $\mu, \zeta, D_r$ should be explored. We find that at $\omega=-1.11/$s and $v_0 = 2\cdot 15.8$ cm/s (factor two to compensate the restoring force from the ring) we find synchronization for example at $\mu=1/2$, $D_r=2$/s and $\zeta=2$/s. This corresponds to a roughly fivefold increase in noise and decrease in self-alignment to parameters calibrated to the single-particle experiments. Here it is important to note, that these are the internal bot parameters, whereas the measurements in the main text are for the augmented unit of bot and housing. A significant deviation has to be expected, as coupled oscillators are known to easily synchronize~\cite{acebron2005}. We show data from a typical run in fig.~\ref{fig:simulation2}. As indicated already in the main text, we find that the bots' inherent chirality prevails without wall interactions and the dominant share of rotational states aligns with their handedness. Using a simple threshold criterion in the center-of-mass-velocity, we can measure the duration of the two motility states. At short times, the details of the classification will matter, but for the tail behavior they are not important. In line with that, we find that the tail behavior is indeed reflective of a Poissonian process with constant rates.
 
\section{Video files}

In addition to the text, we provide the following video files that display the described phenomenology.

\noindent \textsc{orientation.mp4} - Simple experiment to test the self-alignment strength. This was similarly performed in ref.~\citenum{baconnier2022}.\\
\textsc{nautilus.mp4} - Exemplary run of augmented bristlebots in  chiral environment.\\
\textsc{triangle.mp4} - Exemplary run of the active solid structure showing two distinct collective motility states. Realizing a prediction of ref.~\citenum{hernandezlopez2024}.\\
\textsc{triangle\_num.mp4} - Exemplary run of the numerical version of a similar experiment. The video is operating at fixed center of mass, but its motion is indicated in the corner.

\bibliographystyle{apsrev4-1} 
\bibliography{references_arxiv.bib}